\documentclass{jfm}

\usepackage{color}
\usepackage{graphicx}
\usepackage{tikz} 
\usepackage{ulem}
\usepackage{amsmath}
\usepackage{amssymb}

\graphicspath{{./}}

\newcommand{\colr}[1]{\textcolor{red}{#1}}
\newcommand{\colb}[1]{\textcolor{blue}{#1}}
\newcommand{\colm}[1]{\textcolor{magenta}{#1}}

\def\ol   {\overline}
\def\de   {\triangle}
\def\wt   {\widetilde}

\newcommand{\mygam}[2]{\gamma^{#1}_{#2}}
\def\gamvv  {{\gamma_{vv} }}
\def\gamvvs  {{\gamma^*_{vv} }}
\def\gamvvp  {{\gamma^+_{vv} }}
\def\gamuv  {{\gamma_{uv} }}
\def\gamuvs  {{\gamma^*_{uv} }}

\def\be     { \begin{equation} }
\def\ee     { \end{equation} }
\def\bea     { \begin{eqnarray} }
\def\eea     { \end{eqnarray} }
\def\bdm     { \begin{displaymath} }
\def\edm     { \end{displaymath} }
\def\bi     { \begin{itemize} }
\def\ei     { \end{itemize} }
\def\ben     { \begin{enumerate} }
\def\een     { \end{enumerate} }

\def\ap   {\approx}

\def\bp { \begin{picture}(1,1) }
\def\ep { \end{picture} }
\def\stau {{\tau}}

\newcommand{\rfig}[1] {figure~\ref{fig:#1}}
\newcommand{\rtab}[1] {table~\ref{tab:#1}}
\newcommand{\rTab}[1] {Table~\ref{tab:#1}}

\def\dilwrms {\theta_{w,rms}}
\def\dilpwrms {\theta^+_{w,rms}}
\def\urms {u_{rms}}
\def\vrms {v_{rms}}
\def\Trms {T_{rms}}
\def\cut {C_{uT}}
\def\yp {y^+}
\def\yst {y^*}
\definecolor{royalblue}{rgb}{0.2539,0.4102,0.8789}
\definecolor{salmon}{rgb}{0.9792,0.5000,0.4453}

\title{Mach number and wall thermal boundary condition effects on near-wall compressible
turbulence}

\shorttitle{Mach and thermal wall condition effects on near-wall turbulence}
\shortauthor{Akanksha Baranwal, Diego A. Donzis and Rodney W. Bowersox}
\author{
  Akanksha Baranwal\aff{1} \corresp{\email{abaranwal03@tamu.edu}}, 
  Diego A. Donzis\aff{1} , 
  and 
  Rodney D. W. Bowersox\aff{1}}
\affiliation{\aff{1}Department of Aerospace Engineering
Texas A\&M University, College Station, Texas 77843, USA}

\date{\today}

\begin{document}
\maketitle

\begin{abstract}

We investigate the effects of thermal boundary conditions and Mach number on turbulence close to walls.  
In particular, we study the 
near-wall asymptotic behavior for adiabatic and pseudo-adiabatic walls, and compare to the asymptotic behavior recently found
near isothermal cold walls (\cite{BDB2022}). 
This is done by analyzing a new large database 
of highly-resolved direct numerical simulations
of turbulent channels with different wall thermal conditions and centerline Mach numbers. 
We observe that the asymptotic
power-law behavior of Reynolds stresses as well as 
heat fluxes
does change with both centerline Mach number and 
thermal-condition at the wall. 
Power-law exponents transition from their analytical
expansion for solenoidal fields to those for non-solenoidal
field as the Mach number is increased, 
though this transition is found to be dependent on the 
thermal boundary conditions. 
The correlation coefficients between velocity and temperature are also found to be affected by these factors.
Consistent with recent proposals
on universal behavior of compressible turbulence,
we find that dilatation
at the wall is the key scaling parameter for this power-law exponents providing a universal 
functional law which can provide a basis 
for general models of near-wall behavior.

\end{abstract}

\section{Introduction}

The detailed dynamics of turbulence near the wall
has first-order effects on 
phenomena such as heat transfer and viscous drag.
When speeds are relatively low,
many aspects of these flows are relatively well understood 
such as scaling laws for mean quantities and Reynolds stresses.
The situation is more challenging at higher speeds where compressibility
effects become important and the physics more involved due to the interaction 
of hydrodynamics with thermodynamics.
Understanding the detail dynamics in such regimes is critical for
accurate predictions and, ultimately, control of these flows.
It is also critical for model development in the context of 
Reydnolds Averaged Navier-Stokes (RANS) approaches which are widely
used in applications.
Substantial effort have been devoted to 
develop RANS models 
for compressible wall-bounded flows, with adiabatic and weekly 
cooled walls \citep{M1992, SA1992, CA2000}. 
However, these models result 
in poor prediction of statistics at high speeds
\citep{RB2006, R2010, ABDH2022} due to the 
lack of an accurate representation of the different physics and flow behavior in different conditions.    
One important difference between different regimes is the wall thermal 
boundary condition (WTBC) which, in general, is modeled as adiabatic at supersonic 
speeds but cold-wall isothermal in hypersonic regimes.
In certain situations, it is also possible to have mixed boundary conditions
which can again alter the flow dynamics.

Direct numerical simulations (DNS) of a number of wall-bounded 
flows, such as channels \citep{CKM1995,HCB1995,FSF2004,MTN2004,GV2014,SCG2017,
YXP2019,YH2020} and flat plate boundary layers (\citep{SD.book.2006,WSKR2018} 
and references therein) 
have been conducted to try to understand  compressibility 
effects on turbulent statistics in high-speed regimes.
Efforts have also been made to study the effects of WTBC
on the scaling of velocity and temperature statistics and the 
relationship between them in high-speed regimes
\citep{HCB1995,MTN2004,TM2006,thesis_mader,DBM2010,SHH2015,HNSC2015,SLGC2011,ZBHS2014,ZDC2018,ZSLX2022}.
Recent studies have investigated the effects of thermal wall condition on 
pressure fluctuations \citep{ZDC2017,ZWLSL2022}, 
kinetic energy transfer (\cite{XWYLC2021}), density and temperature 
resolvent mode shapes (\cite{aiaa_HSB2020}) highlighting WTBC effects on turbulent processes and structures.
Several studies 
focused on finding scaling laws and others on using these scaling
laws to collapse first and
second order statistics in high-speed regimes for different 
flow conditions and different WTBCs \citep{BBHC2008,ZBHLS2012,TL2016,PPBP2015,VIPL2020,GFM2021}.
These WTBCs can be broadly characterized as isothermal (constant
temperature) and
isoflux (constant heat flux) conditions. 
For the former, studies have been conducted to investigate 
the effect of wall temperature,
and for the latter, the effect of varying rate of heat transfer.
Some studies have also used the so-called pseudo-adiabatic wall, 
a constant wall 
temperature (based on the recovery factor) whose value is such that 
the mean heat transfer to the wall
vanishes, mimicking an adiabatic boundary. 
Some of the studies mentioned above
(\cite{SHH2015,WSKR2018,ZSLX2022}) found that 
variation in turbulent statistics (e.g., mean velocity, mean temperature,  
Reynolds stresses) are not due to changes 
in the WTBC itself (i.e., change from isothermal to isoflux), 
but instead due to change in the heat transfer at the wall. 
However, direct effects of changing the boundary condition from isothermal to 
isoflux were observed on temperature fluctuation statistics, 
e.g.\ temperature fluxes in
the near-wall region and these effects extended beyond the viscous sublayer.
Another important observation was the change in asymptotic behavior 
of turbulent heat fluxes for different WTBCs.
   
From a fundamental and a modeling perspective, 
it is crucial to understand the precise asymptotic 
behavior of turbulence close to the wall. 
Indeed, accurate predictions necessitates models to 
satisfy the correct asymptotic scaling laws
\citep{LS1990,SLZH1991,SZS1991,ZSSL1992,
durbin1993,SSZ1993,SGS1998,GPM1991,bowersox2009,AWGBM2022}, 
and thus many studies have reported 
the asymptotic behavior of turbulent fluxes for both incompressible and compressible
flows and under different WTBCs
(\cite{MTN2004,SHH2015,HNSC2015,ZSLX2022,LSBH2009}).
All these studies compared 
their data to the theoritical asymptotes obtained from Taylor series 
expansions in wall-normal direction and found good agreement.       
However, none of these studies examined well-resolved wall asymptotes with a 
systematic variation of Mach number for different WTBCs. 

We have recently 
conducted DNS of turbulent channels 
with finer near-wall resolution than the standard in the literature
to capture true asymptotic behavior (\cite{BDB2022}).
In that study, which was done with cooled isothermal walls,
we systematically varied the centerline Mach number 
from $M \gtrsim 0.2$ (virtually incompressible) to $M \lesssim 2.2$.
We showed that
turbulent stresses and wall-normal heat flux comprising at least 
one wall-normal velocity component do not collapse when the Mach number
was changed as suggested by widely used scaling laws which, thus, undermines
Morkovin's hypothesis.
In particular, due to the extremely high wall resolution,
we were able to unveil a new region very close to the wall where
power-law scaling exponents were found to differ from 
theoretical asymptotes and, furthermore, depend on Mach number. 
Previous studies at the standard resolution are not able to capture 
this region.   
We have also found that increasing the centerline Mach number 
resulted in enhanced levels of dilatation motions at the wall which is 
the key factor to understand changes in the power-law asymptotes close to the wall. 
Dilatational levels at the wall were also found to be affected by WTBCs in
boundary layers
\citep{XWYLC2021,ZWLSL2022}. 
\cite{ZWLSL2022} further found that 
wall cooling effects on dilatation depends also on the Mach number. 

These complex dependencies on both WTBCs and Mach number is the 
motivation behind the present work.
In particular we investigate, for the first time, the asymptotic
behavior of various turbulent stresses and heat fluxes at different
Mach numbers and for different WTBCs. This systematic investigation 
is possible due to extremely well resolved turbulent channels with 
centerline Mach number ranging from 0.2 and 2.2 with isothermal,
adiabatic and pseudo-adiabatic walls. The new adiabatic and 
pseudo-adiabatic results
complement the isothermal data in \cite{BDB2022}.
This is also relevant in the context of classical 
scaling laws based on Morkovin's hypothesis which 
are more effective at collapsing statistics when the walls are adiabatic 
or weakly-cooled, than when they are isothermal in which case there is significant 
wall cooling. 
Adiabatic walls, thus, possess the additional 
advantage of isolating the effects of Mach number 
from wall cooling and provide a more direct way to assess the effects 
of Mach number in isolation 
and the validity of Morkovin's hypothesis on the asymptotic 
scaling of turbulent statistics. 

The rest of the paper is organized as follows. We first present 
the numerical method, configuration, and DNS database.
Then, we present results on the asymptotic behavior of 
Reynolds stresses and their dependency on centerline Mach number and WTBCs.
This analysis is 
then extended to temperature fluctuations 
and heat fluxes. We conclude with a summary and some remarks on 
the implications of the results presented here.

\section{Numerical Method} \label{sec:DNS}
We perform direct numerical simulations of the equations governing mass,
momentum, and energy conservation
for a compressible channel flow. 
The equations are discretized on a uniform mesh in the streamwise ($x$) and 
spanwise ($z$) directions.
In the wall-normal ($y$) direction,
the grid is clustered close to the wall using a hyperbolic tangent function.
We use sixth-order compact schemes to compute spatial
derivatives in the $x$ and $z$ directions.
For the $y$ direction, we utlize the sixth-order compact
scheme in interior points and the order is reduced 
to fourth and third 
at the last two grid points in the domain. 
The variables are marched in time using a third-order low-storage
Runge-Kutta scheme.  
More details on simulations can be found in \cite{BDB2022} where we
also present 
detailed grid convergence studies and  
validations against other DNS databases in the literature 
\citep[e.g.][]{CKM1995}.
The simulations presented here satisfy those resolution criteria
which are summarized in \rtab{dns1}.

Periodic boundary conditions are used in the 
streamwise and spanwise directions. 
At the walls, we apply no-slip 
boundary conditions for all velocity components.
The boundary condition for pressure is obtained by evaluating 
the momentum equation in the normal direction at the wall
which was found to have a greater numerical stability than the commonly used
zero-pressure gradient \citep{BDB2022}.
In all the simulations presented here, 
the bottom wall ($y=0$) is isothermal with $T=300$.
For the top wall,
three 
different thermal boundary conditions are investigated, namely,
isothermal, adiabatic and pseudo-adiabatic 
cases denoted by I, A and PA respectively. 
For isothermal cases, the top wall is kept 
at the same temperature as the bottom wall ($T=300$). 
These simulations, which 
were studied in our previous study (\cite{BDB2022}), act as base case 
to compare with other thermal wall conditions.
For adiabatic cases, we specify zero temperature gradient 
at the top wall. 
This 
approach with mixed boundary conditions in a channel have been
used before \citep{MTN2004,TM2006,ZSLX2022,LC2022,aiaa_BDB2023}.
Finally, the pseudo-adiabatic case consists of imposing 
an isothermal boundary condition 
at the average temperature obtained  
from the adiabatic simulation 
with all other flow parameters kept the same as in the adiabatic case.

Following standard notation, 
the bulk, wall and centerline values of a variable $f$ are
denoted by $f_b$, $f_w$ and $f_c$, respectively.
Reynolds and Favre decompositions are
denoted by $\ol{q}+q'$ and $\wt{q}+q''$, respectively.
The averages in these decompositions are
taken along the homogeneous directions (i.e.\ $x$-$z$ planes) and time. As done
in \cite{BDB2022}, 
snapshots of all fields 
are saved at time intervals of $5h/\ol{u_b}$ for all simulations 
, where $h$ is the channel half width and 
$u$ is the streamwise velocity component. 
This time scale ($h/\ol{u_b}$) is commensurate with the eddy-turnover time
of the turbulence in the center of the channel and thus 
representative of the largest turbulent structures.  
Our temporal averages involved 25 snapshots for velocity, density and 
temperature fields. 

\begin{figure}
\centering
\includegraphics[width=.42\textwidth]{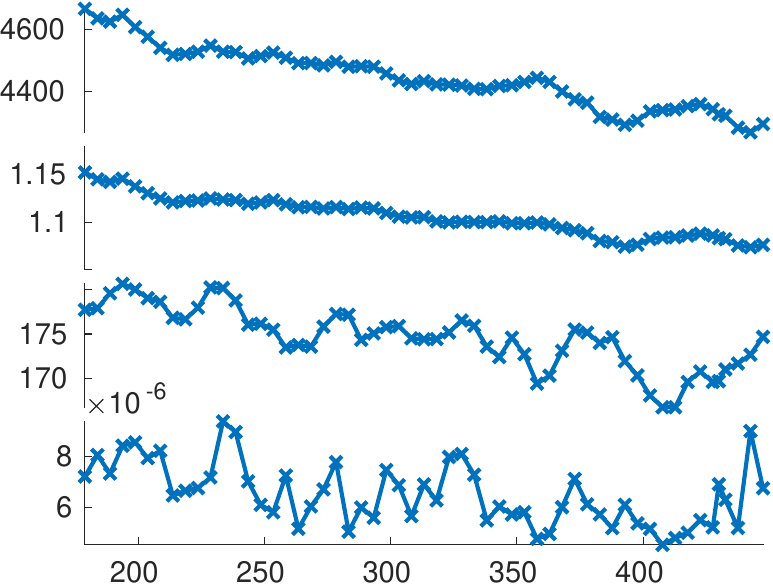}
\begin{picture}(0,0)
\put(-183,120){ (a)}
\put(-172,98){\rotatebox{90}{{$Re_c$}}}
\put(-170,70){\rotatebox{90}{{$M_c$}}}
\put(-170,38){\rotatebox{90}{ {$Re_{\tau}$}}}
\put(-170,5){\rotatebox{90}{{$\dilwrms^+$}}}
\put(-80,-15){{$tu_b/h$}}
\end{picture}
\hspace{8 mm}
\vspace{6 mm} 
\includegraphics[width=.48\textwidth]{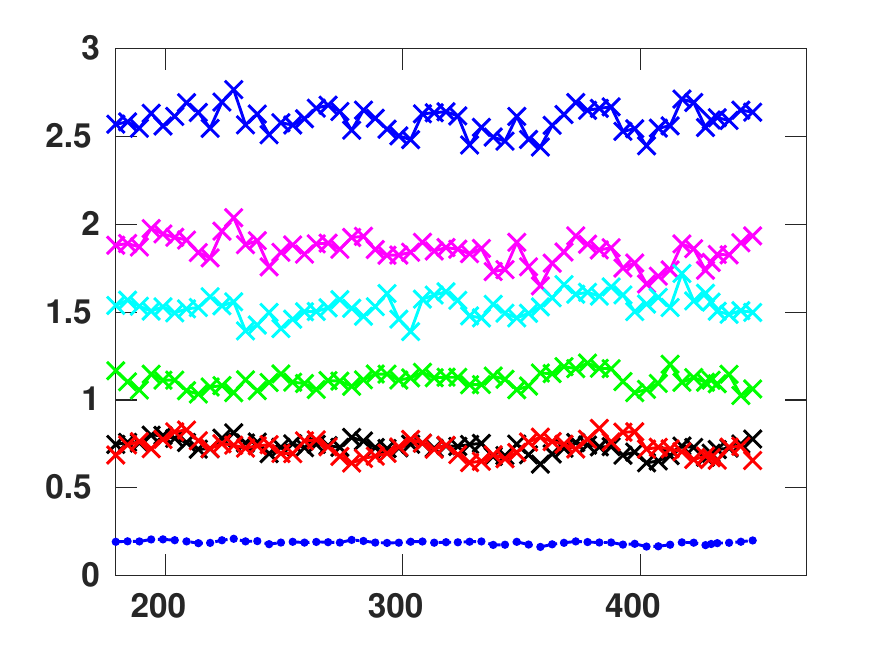}
\begin{picture}(0,0)
\put(-200,120){(b)}
\put(-195,50){\rotatebox{90}{ {$u_{rms}/u_{\tau}$}}}
\put(-103,-15){ {$tu_b/h$}}
\end{picture}
\includegraphics[width=.48\textwidth]{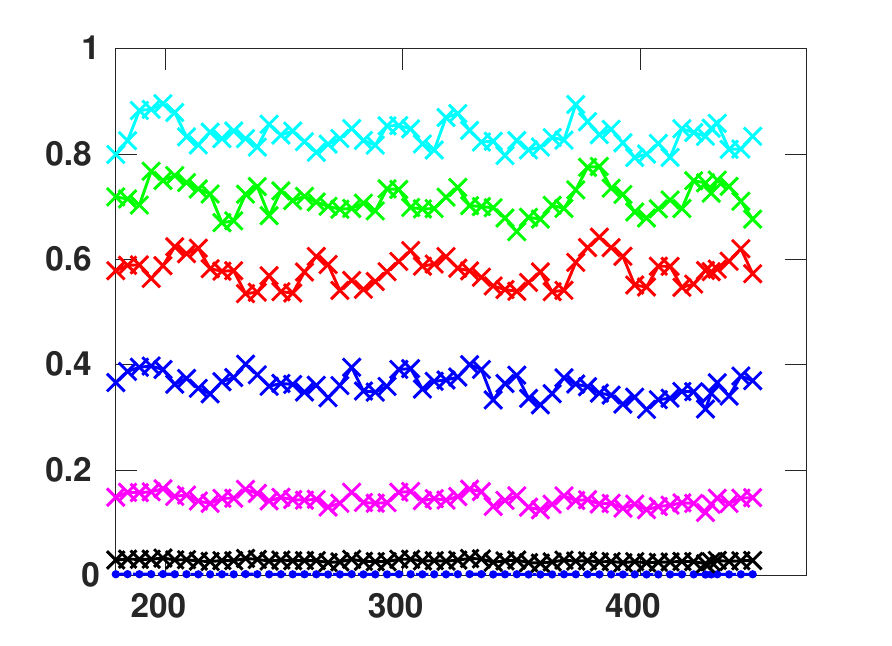}
\begin{picture}(0,0)
\put(-200, 120){(c)}
\put(-195,50){\rotatebox{90}{ {$v_{rms}/u_{\tau}$}}}
\put(-103,-15){{$tu_b/h$}}
\end{picture}
\includegraphics[width=.48\textwidth]{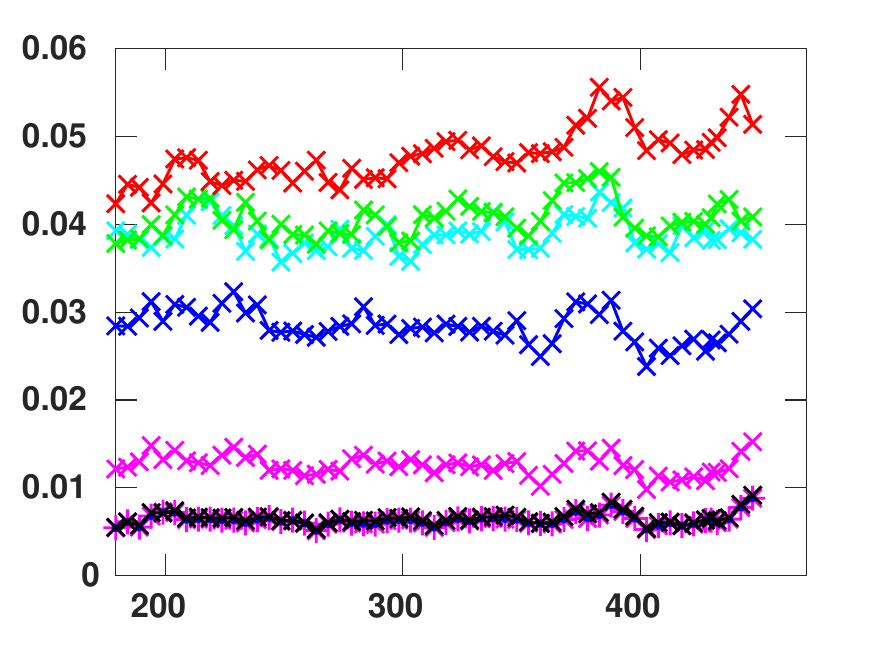}
\begin{picture}(0,0)
\put(-200, 120){(d)}
\put(-200,50){\rotatebox{90}{ {$T_{rms}/\ol{T}$}}}
\put(-103,-15){{$tu_b/h$}}
\end{picture}
\vskip 6 mm
\caption{Time evolution of (a) $Re_{c}$, $M_c$, $Re_{\tau}$, $\dilwrms^+$, 
(b) $u_{rms}/u_{\tau}$, (c) $v_{rms}/u_{\tau}$, (d) $T_{rms}/\ol{T}$  
for $M_b \approx 1.2$,  (\colm{**** - $y^+=0$}), (\colb{....- $y^+=0.5$}), ({xxxx - $y^+=2.0$}),
	(\colm{xxxx - $y^+=5.6$}), (\colb{xxxx - $y^+=11$}), (\textcolor{cyan}{xxxx - $y^+=54$}), 
	(\textcolor{green}{xxxx - $y^+=100$}), (\textcolor{red}{xxxx - $y^+=173$}). } 
\label{fig:tseries}
\end{figure}
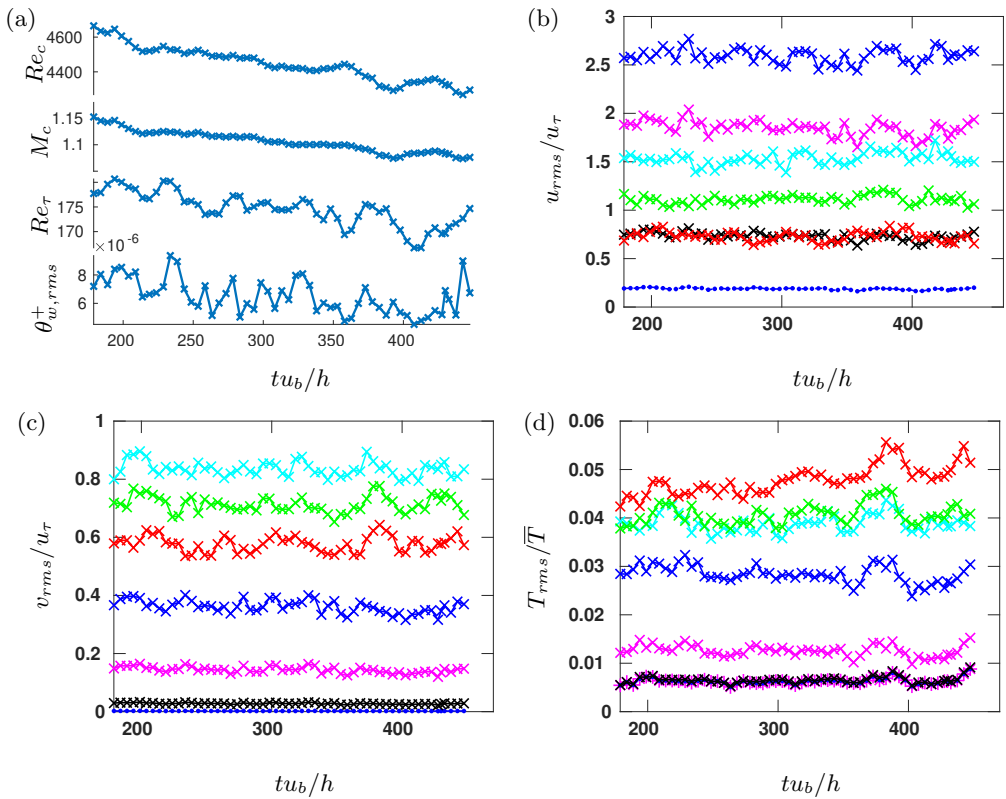

Consider a forced, periodic channel with Dirichlet boundary 
conditions for temperature at the walls, that is isothermal walls. 
If initialized with zero velocity and constant temperature, the flow will 
accelerate and develop velocity gradients that lead to 
viscous dissipation. This leads to an increase in temperature 
inside the channel which is higher than that imposed at the walls.
Because of the thermal gradient that forms at the wall, there is a flux of energy 
from the fluid to the wall and the flow eventually reaches a
statistically steady state where the rate of production of internal energy 
due to viscous dissipation is compensated by the energy transfer through the wall.
If, on the other hand, 
we apply a Neumann boundary condition for temperature at the wall, 
in particular zero temperature gradient, then the heat transfer to the walls 
is identically zero. 
In this case, 
the increase in temperature 
due to dissipation maintained by the forcing in the momentum equation 
is not balanced by heat flux through the walls.
Therefore, the internal energy in the channel increases continuously leading to 
a time-dependent mean thermodynamic state. 
Alternatively, one can apply a (cold) isothermal condition 
to one wall and an adiabatic condition the other wall.
This allows for heat transfer through one wall and 
results in a decreased rate of change of mean thermodynamic 
parameters. In this case, the flow also achieves a 
{\it pseudo-steady state} where statistics 
(at least to second order) are in a statistically steady state when normalized
by their corresponding (slowly varying) means.
This can be seen in \rfig{tseries}(b)(c)(d) where we show the 
temporal evolution of the root-mean-square (r.m.s.)
of several variables normalized by their respective time-varying means
for $M_b\approx 1.2$ and very long simulation time ($\approx$ $200h/u_b$).
While global quantities (Reynolds and Mach numbers 
in panel (a)) are seen to decrease slowly, normalized 
fluctuations statistics are virtually in a statistical steady state.
This is, in fact, consistent with observation in forced
isotropic flows \citep{KO1990}.
We do note that there seems to be a (very weak)
increase in $T_{rms}$ 
at the centerline ($y^+ = 173$, red symbols). 
Because our 
interest lies close to the wall, we have verified this trend very far 
from the wall is not a concern in this study.
The normalized r.m.s. dilatation at the wall, as shown in \rfig{tseries} (a), 
$\dilwrms^+ = \ol{(\partial v'/\partial y)_w^2}\nu_w^2/u_\tau^4$ 
is another quantity of interest which also exhibits a steady-state behavior. 
We take advantage of this pseudo-steady state to find averages 
over the simulation time. The statistics below are based on this 
averaging.

The friction Reynolds numbers based on wall quantities and the friction 
Reynolds numbers based on centerline viscosity and density, are defined as
$Re_\stau = \ol{\rho_w}  u_\stau h / \ol{\mu_w}$
and 
$Re_\stau^* \equiv  \ol{\rho_c} {(\tau_w/ \ol{\rho_c}})^{1/2} h / \ol{\mu_c}$ 
respectively, 
with $u_\stau \equiv  \sqrt{\tau_w/\ol{\rho_w}}$ being the friction velocity.
The centerline Reynolds number and centerline Mach numbers are 
$Re_c \equiv \ol{\rho_c}$ $\ol{u_c} h / \ol{\mu_c}$ 
and $M_c \equiv \ol{u_c}/ \sqrt{\gamma R\ol{T_c}}$, respectively. 
Our domain has dimensions $4\pi h \times 2h \times 4\pi/3h $ for all
our simulations.  
This is larger than widely used in literateure (\citep[e.g.][]{TL2016,YXP2019}).
Finally, as a direct assessment of boundary conditions effects 
on the quantities studied here, 
we have run additional simulations with a domain which is 20\% shorter 
and confirmed that the near-wall scaling laws are unaffected.
\rTab{dns1} summarizes the important parameters for the DNS database
used here. 

\begin{table}
\begin{center}
\begin{tabular}{p{2.75cm}p{0.75cm}p{0.75cm}p{0.75cm}p{0.65cm}p{0.95cm}
p{0.95cm}p{0.85cm}p{0.75cm}c}
        \hline
 		\hskip 0.8 cm Wall & $ M_c$&$Re_c$&$Re_{\tau}$& $Re_{\tau}^*$ &  $\de y_{min}^+$&  $\de y_{max}^+$&  $\de x^+$& 	  $\de z^+$  &  Line style\\
        \hline
          \hfil Isothermal & 0.23 & 5692 & 295 &{293} & {0.08} & 2.9 & 14.5 & 4.8 & \tikz\draw [red,thick] (0,0) -- (0.7,0);\\
         \hfil Adiabatic & 0.23 & 5684 & 296 &{292} & {0.08} & 2.9 & 14.5 & 4.8 & \tikz\draw [dashdotted,red,thick] (0,0) -- (0.7,0);\\ 
         \hfil Isothermal & 0.35 & 5638 & 294 &{289} & {0.08} & 2.9 & 14.4 & 4.8 & \tikz\draw [brown,thick] (0,0) -- (0.7,0);\\  
         \hfil Isothermal & 0.46 & 5582 & 294 &{286} & {0.08} & 2.9 & 14.4 & 4.8 & \tikz\draw [orange,thick] (0,0) -- (0.7,0); \\ 
         \hfil Isothermal & 0.57 & 5476 & 293 & 281 & {0.05} & 3.2 & 14.4 & 4.8 & \tikz\draw [cyan,thick] (0,0) -- (0.7,0);\\
         \hfil Adiabatic & 0.57 & 5476 & 293 & 281 & {0.05} & 3.2 & 14.4 & 4.8 & \tikz\draw [dashdotted,cyan,thick] (0,0) -- (0.7,0);\\ 
        \hfil Isothermal & 0.68 & 5498 & 301 &{283} & {0.05} & 3.3 & 14.8 & 4.9 &\tikz\draw [teal,thick] (0,0) -- (0.7,0);\\ 
         \hfil Isothermal & 0.89 & 5371 & 307 &{276} & {0.05} & 3.4 & 15.1 & 5.0 &\tikz\draw [blue,thick] (0,0) -- (0.7,0);\\ 
         \hfil Adiabatic & 0.84 & 5099 & 225 &{260} & {0.05} & 2.2 & 11.0 & 3.6 &\tikz\draw [dashdotted,blue,thick] (0,0) -- (0.7,0);\\ 
         \hfil Isothermal & 1.26 & 5022 & 325 &{259} & {0.05} & 3.6 & 15.9 & 5.3 &\tikz\draw [black,thick] (0,0) -- (0.7,0);\\ 
    \hfil Adiabatic & 1.12 & 4513 & 177 &{226} & {0.05} & 1.7 & 8.7 & 2.9 &\tikz\draw [dashdotted,black,thick] (0,0) -- (0.7,0);\\ 
                 \hfil \footnotesize Pseudo-adiabatic & 1.12 & 4306 & 179 &{220} & {0.05} & 1.7 & 8.8 & 2.9 &\tikz\draw [dashed,black,thick] (0,0) -- (0.7,0);\\ 
        \hfil Isothermal & 1.50 & 5489 & 393 &{277} & {0.10} & 4.0 & 19.3 & 6.4 &\tikz\draw [gray,thick] (0,0) -- (0.7,0);\\ 
        \hfil Isothermal & 1.98 & 5631 & 572 &{279} & {0.10} & 6.2 & 14.0 & 4.7 &\tikz\draw [magenta,thick] (0,0) -- (0.7,0);\\
                \hfil Adiabatic & 1.9 & 5092 & 138 &{236} & {0.05} & 1.0 & 3.4 & 1.0 &\tikz\draw [dashdotted,magenta,thick] (0,0) -- (0.7,0);\\
        \hfil Isothermal & 2.22 & 5666 & 745 &{273} & {0.09} & 8.8 & 14.8 & 6.1 &\tikz\draw [purple,thick] (0,0) -- (0.7,0);\\
        
\end{tabular}
\caption{Details of flow conditions and grid resolutions}
\label{tab:dns1}
\end{center}
\end{table}

In subsequent sections, 
we investigate 
various statistics
near isothermal (solid lines) and adiabatic walls (dash-dotted lines) for three different 
centerline Mach numbers, $M_c \approx 0.23$ (red), 
$M_c \approx 1.2$ (black) and $M_c \approx 1.9$ (magenta) and near pseudo-adiabatic walls (dashed lines)
for $M_c \approx 1.2$ using our DNS database.  
The adiabatic  and 
pseudo-adiabatic results are taken from the upper halves of channel from the A and PA simulations 
respectively where bottom walls are isothermal. 
 The isothermal  case throughout the work refers to simulations where both walls are isothermal, unless specifically noted otherwise. We note that 
 wall quantities for a particular case
refer to the statistics at the wall with that particular thermal boundary condition
(e.g. for pseudo-adiabatic case, wall quantities refer 
to statistics at pseudo-adiabatic wall).

\def\picw {0.43}
\def\lela {\tikz\draw [red,ultra thin] (0,0) -- (0.4,0)}
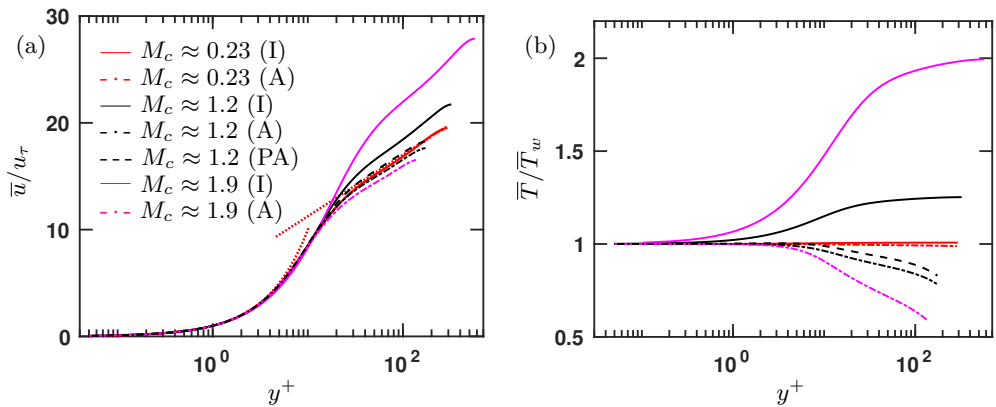
\begin{figure} 
\centering
\includegraphics[width=\picw\textwidth]{ 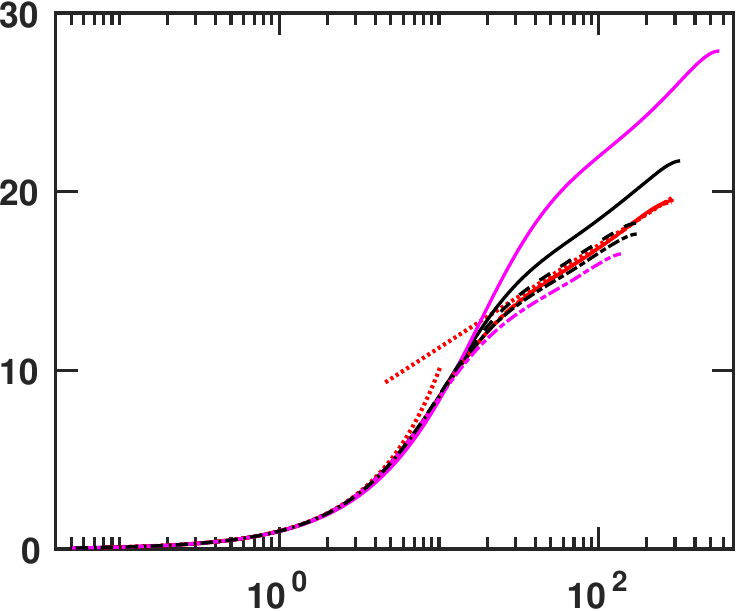}
\begin{picture}(0,0)
\put(-183, 120){ (a)}
    \put(-183,63){\rotatebox{90}{ {$\ol{u}/u_{\stau}$}}}
    \put(-148,120){\tikz\draw [red,ultra thin] (0,0) -- (0.4,0); }
    \put(-133,118) {$  M_c \approx 0.23$ (I)}
    \put(-148,108){\colr{-\,$\cdot$\,-} $M_c \approx 0.23$ (A)}
       \put(-148,100){\tikz\draw [black,ultra thin] (0,0) -- (0.4,0); }
     \put(-133,98){$M_c \approx 1.2$ (I)}
    \put(-148,88){{-\,$\cdot$\,-} $M_c \approx 1.2$ (A)}
    \put(-148,78){{-\,-\,-} $M_c \approx 1.2$ (PA)}
    \put(-148,71){\tikz\draw [magenta,ultra thin] (0,0) -- (0.4,0);}
     \put(-133,68){$  M_c \approx 1.9$ (I)}
    \put(-148,58){\colm{-\,$\cdot$\,-} $M_c \approx 1.9$ (A)}
 \put(-85, -10){{$y^+$}}
\end{picture}
 \hspace{6 mm}
\vspace{6 mm}
 \includegraphics[width=\picw\textwidth]{ 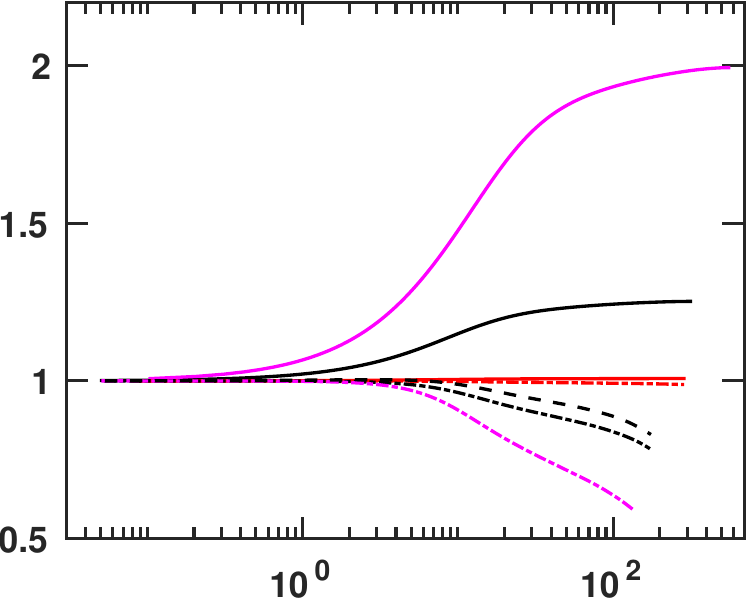}
 \begin{picture}(0,0)
 \put(-183, 120){ (b)}
 \put(-183,65){\rotatebox{90}{$\ol{T}/\ol{T}_w$}}
 \put(-85, -10){{$y^+$}}
 \end{picture}
  \caption{(a) Streamwise mean velocity normalized by the friction velocity 
(b) Mean temperature normalized by the mean wall temperature plotted against 
wall-normal coordinate in viscous units  
for isothermal 
        (---), adiabatic (-\,$\cdot$\,-) and pseudo-adiabatic 
(-\,-\,-) cases. 
	Red, black and magenta correspond to $M_c \approx 0.23$, 
$M_c \approx 1.2$ and $M_c \approx 1.9$ respectively.
Dotted red lines represent viscous and log layer scalings.}
\label{fig:umean3}
\end{figure}

\section{First-order statistics}

The mean streamwise velocity normalized by the friction velocity 
and the mean temperature normalized by the wall temperature are
shown in 
\rfig{umean3} (a) and (b) respectively. 
Consistent with the literature, wall-normalization 
($u_{\stau}$) 
performs well in collapsing velocity for all presented cases in the
viscous sublayer and the majority of the buffer layer ($y^+ \lesssim 20$)
as shown in \rfig{umean3} (a).
In the log-law region, such collapse is not observed and 
Mach number and WTBC
effects exist on the mean velocity. In particular, Mach number effects 
are more pronounced in isothermal than in adiabatic cases
due to the increased heat transfer to the wall as 
the Mach number increases for the former. 
 In \rfig{umean3} (b) we find that, for isothermal cases, 
 wall cooling leads to a temperature inside the channel which is higher than
the wall temperature with a maximum  
 at the centerline. 
This maximum temperature along with the wall cooling rate 
 increase with the Mach number. 
 For adiabatic cases, 
because of the zero heat flux at the top wall, there is a rise of 
 temperature across the channel, with the maximum temperature at 
 the upper wall. 
 The temperature gradient for adiabatic and pseudo-adiabatic 
 cases is not zero at the centerline, which may result in non-zero 
 temperature fluxes at the channel half-width, an effect that 
 will be discussed later. 
 
\def\picw {0.295}
\begin{figure} 
\centering
 \includegraphics[width=\picw\textwidth]{ 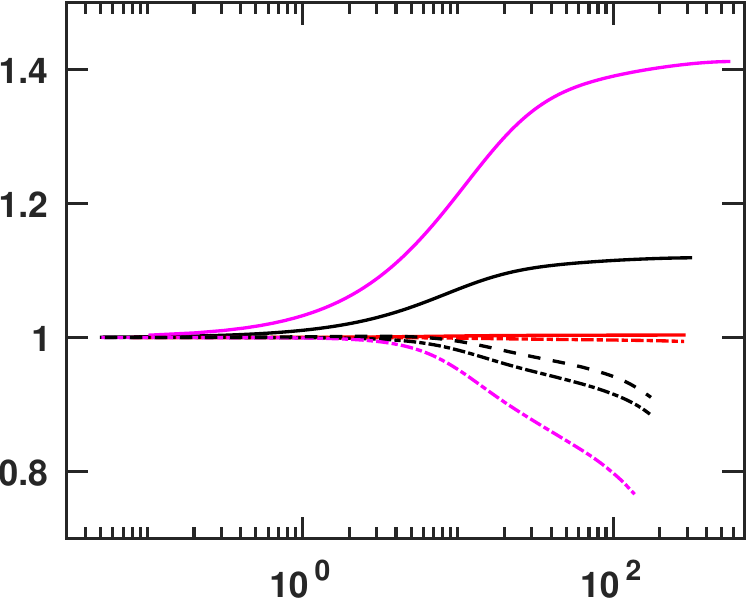}
 \begin{picture}(0,0)
 \put(-135, 85){ (a)}
 \put(-130,45){\rotatebox{90}{$\ol{\mu}/\ol{\mu}_w$}}
  \put(-228,80){\rotatebox{90}{ {$\ol{u}/u_{\stau}$}}}
  \put(-60, -10){$y^+$}
\end{picture}
\hspace{2 mm}
\vspace{6 mm}
\includegraphics[width=0.31\textwidth]{ 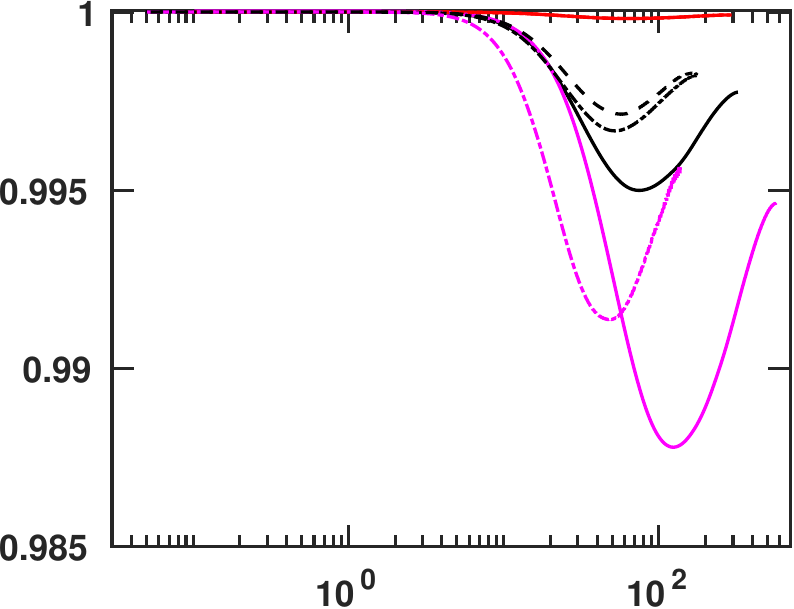}
\begin{picture}(0,0)
\put(-135, 85){ (b)}
 \put(-135,45){\rotatebox{90}{$\ol{p}/\ol{p}_w$}}
  \put(-60, -10){$y^+$}
    \put(-104,80){\tikz\draw [red,ultra thin] (0,0) -- (0.4,0); }
    \put(-90,78) {\tiny {$  M_c \approx 0.23$ (I)}}
     \put(-104,68){\colr{-\,$\cdot$\,-} \tiny{$M_c \approx 0.23$ (A)}}
         \put(-104,58){\tikz\draw [black,ultra thin] (0,0) -- (0.4,0); }
       \put(-90,56){\tiny {$M_c \approx 1.2$ (I)}}
     \put(-104,48){{-\,$\cdot$\,-} \tiny{$M_c \approx 1.2$ (A)}}
     \put(-104,38){{-\,-\,-} \tiny{$M_c \approx 1.2$ (PA)}}
      \put(-104,30){\tikz\draw [magenta,ultra thin] (0,0) -- (0.4,0);}
       \put(-90,28){\tiny {$  M_c \approx 1.9$ (I)}}
      \put(-104,18){\colm{-\,$\cdot$\,-} \tiny{$M_c \approx 1.9$ (A)}} 
\end{picture}
\hspace{3 mm}
\includegraphics[width=0.29\textwidth]{ 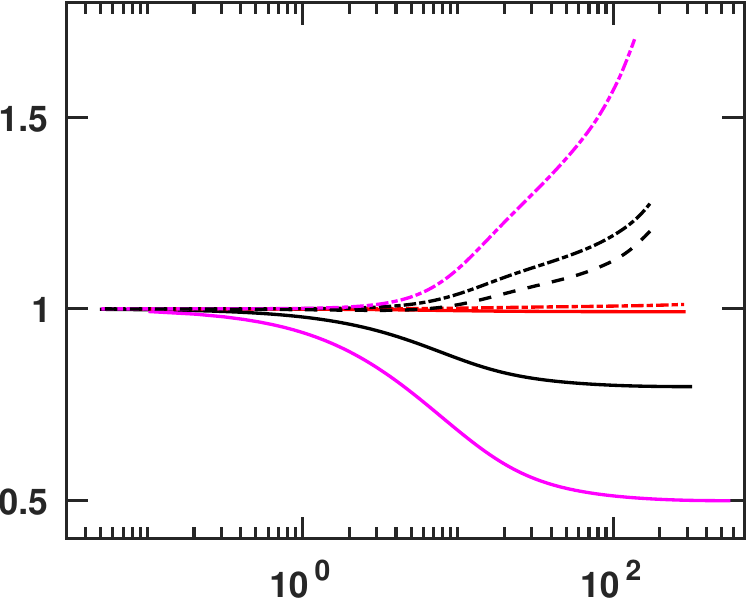}
\begin{picture}(0,0)
\put(-130, 83){ (c)}
 \put(-125,45){\rotatebox{90}{$\ol{\rho}/\ol{\rho}_w$}}
  \put(-60, -10){$y^+$}
\end{picture}
 \includegraphics[width=\picw\textwidth]{ 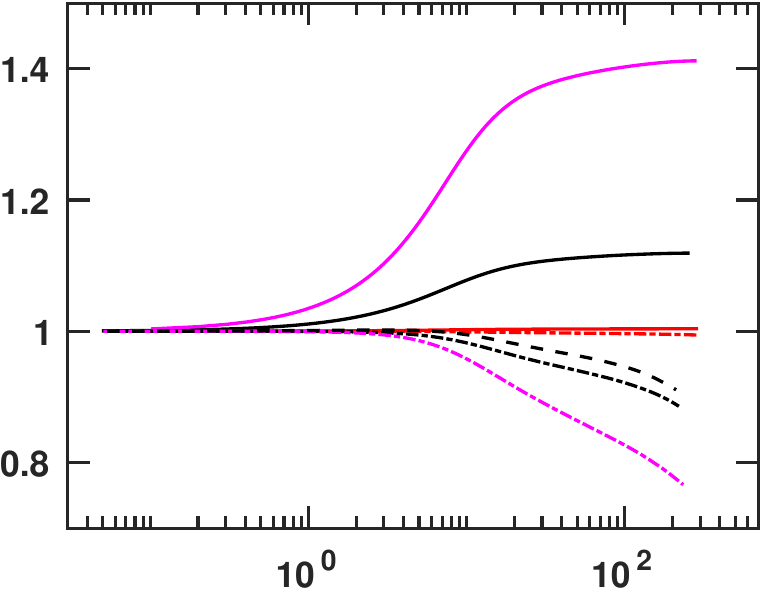}
 \begin{picture}(0,0)
 \put(-135, 85){ (d)}
 \put(-130,45){\rotatebox{90}{$\ol{\mu}/\ol{\mu}_w$}}
  \put(-60, -10){$y^*$}
 \end{picture}
\hspace{2 mm}
\vspace{6 mm}
\includegraphics[width=0.31\textwidth]{ 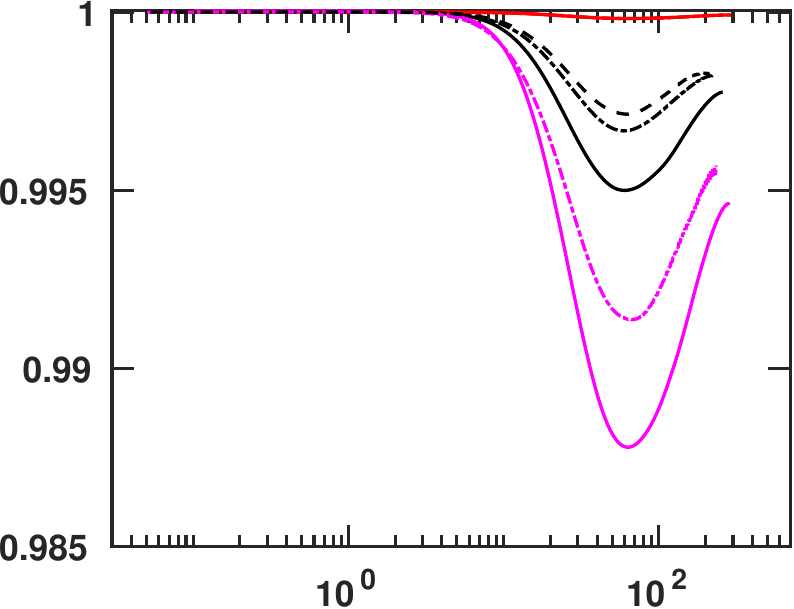}
\begin{picture}(0,0)
\put(-135, 85){ (e)}
 \put(-135,45){\rotatebox{90}{$\ol{p}/\ol{p}_w$}}
  \put(-60, -10){$y^*$}
\end{picture}
\hspace{4 mm}
\includegraphics[width=0.295\textwidth]{ 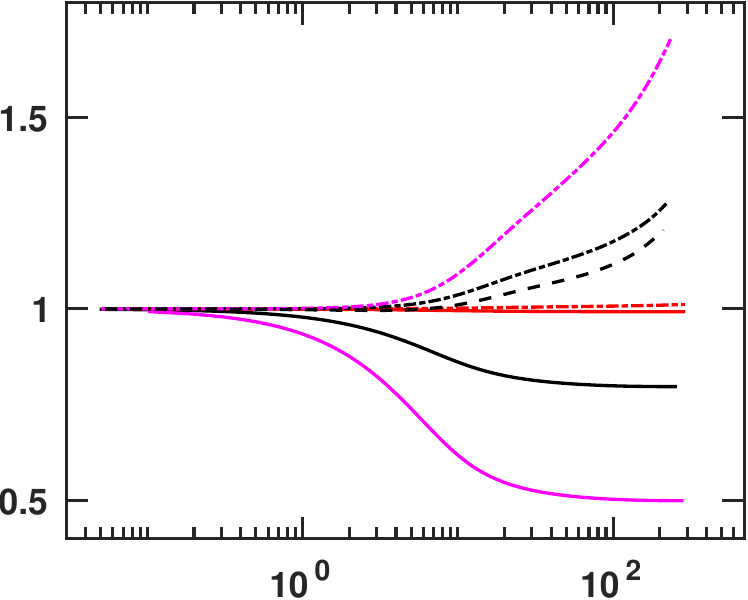}
\begin{picture}(0,0)
\put(-135, 85){ (f)}
 \put(-127,45){\rotatebox{90}{$\ol{\rho}/\ol{\rho}_w$}}
  \put(-60, -10){$y^*$}
\end{picture}
  \caption{Mean viscosity (a,d), pressure (b,e),
  and density (c,f) normalized by their corresponding wall values and 
plotted versus wall-normal coordinate in viscous units (a-c)  
 and semi-local units (d-f)
for  isothermal,
(---) adiabatic (-\,$\cdot$\,-) and pseudo-adiabatic 
(-\,-\,-) cases. 
	Red, black and magenta correspond to $M_c \approx 0.23$, 
$M_c \approx 1.2$ and $M_c \approx 1.9$ respectively.} 
\label{fig:thmean}
\end{figure}
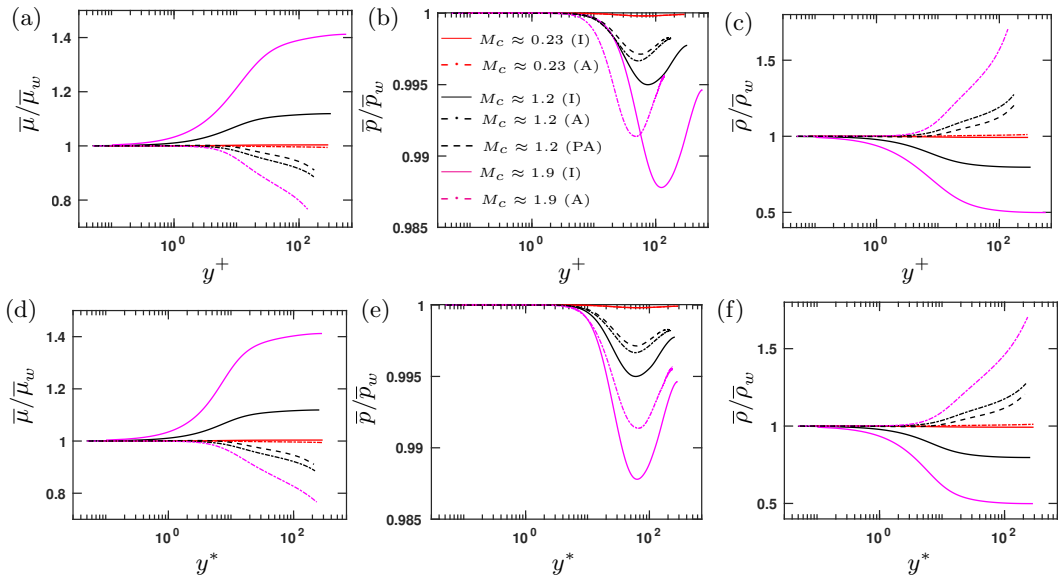

The mean viscosity, mean pressure  and mean density are shown in
\rfig{thmean} (a-f) against wall-normalized (a-c) and semi-local (d-f) wall-normal coordinates. 
 On comparing \rfig{thmean}  (a) with (d), (b) with (e), and (c) 
 with (f), we observe
 that some features become independent of Mach number
 or WTBC when the statistics are plotted 
 against the semi-local wall-normal coordinate 
 $\yst\equiv  \ol{\rho} {(\tau_w/ \ol{\rho}})^{1/2} y / \ol{\mu}$
 as opposed to $\yp$.
 For example, 
in \rfig{thmean} (e)
 we see 
 that pressure start decreasing significantly 
 only at $\yst \approx 5$, reaching a
 minimum at $\yst \approx 65$ for all $M_c$, and then increasing towards the
 channel centerline. 
 Similar observations can be made for viscosity and density for isothermal
 cases (\rfig{thmean} (d, f)).
 As expected, the mean viscosity 
 follows a similar trend as the mean 
 temperature (\rfig{umean3} (b)).
 Pressure is relatively constant across the channel with a small 
 dip outside of the viscous sublayer which increases with $M_c$ to 
 about 1.5\% at the highest Mach number shown ($M_c\approx 1.9$).
In \rfig{thmean} (c) and (f) we show the mean density 
normalized by the mean density at the wall. 
Because the mean pressure is roughly 
constant across the channel, the mean density is inversely 
proportional to the 
mean temperature which is what we observe in these plots.

 We can also see opposite trends depending on WTBC.
 For isothermal cases, the density decreases as one moves 
 away from the wall or when the Mach number increases. 
 For adiabatic and pseudo-adiabatic cases, on the other hand, 
 the density increases as one moves away from the wall or 
 when the Mach number decreases.
 A result of these trends is that,
 close to the wall, the density gradients are 
 higher for isothermal cases at higher Mach numbers
 indicating that the statistics will change more rapidly from their wall values 
 in the near-wall region
 when the level of compressibility and heat transfer to the wall are increased.

\begin{figure} 
\centering
\includegraphics[width=.55\textwidth]{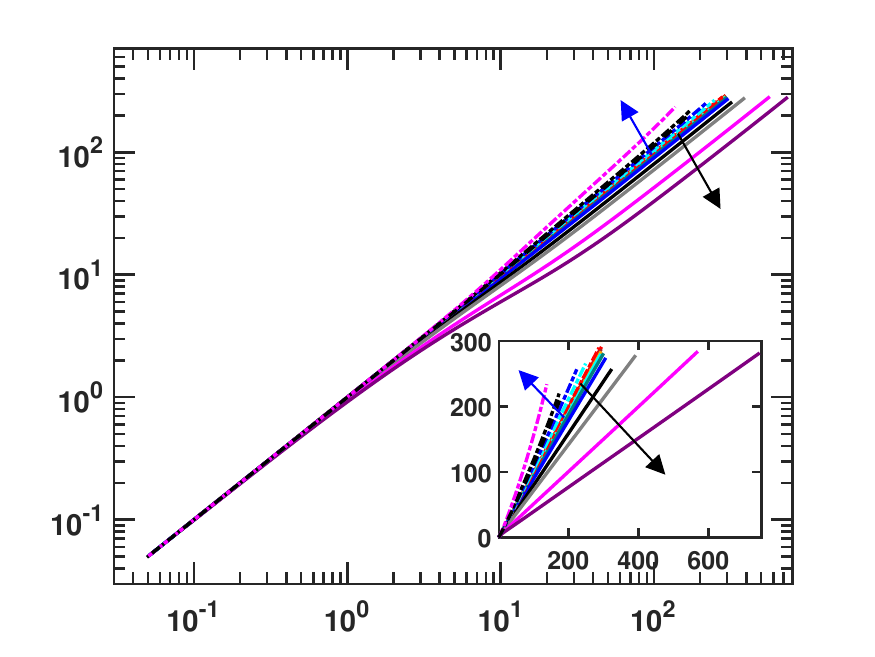}
\begin{picture}(0,0)
   \put(-215,75){\rotatebox{90}{{$y^*$}}}
 \put(-110, -9.5){ {$y^+$}}
 \put(-117,137){\scriptsize{(Adiabatic)}}
 \put(-75,137){ $M_c$}
  \put(-70,92){\scriptsize{(Isothermal)}}
   \put(-45,100){ $M_c$}
         \put(-96,70){\scriptsize { $M_c$}}
   \put(-55,38){\scriptsize{ $M_c$}}
\end{picture}
\vspace{6 mm}
 \caption{ Wall-normal coordinate in semi-local units versus wall-normal coordinate 
in viscous units
for  isothermal (---),
 and adiabatic (-\,$\cdot$\,-)  cases. 
Colors as in \rtab{dns1}. Black and blue arrows 
indicate increase in $M_c$ for isothermal and
adiabatic cases respectively.} 
\label{fig:semilocal}
\end{figure}
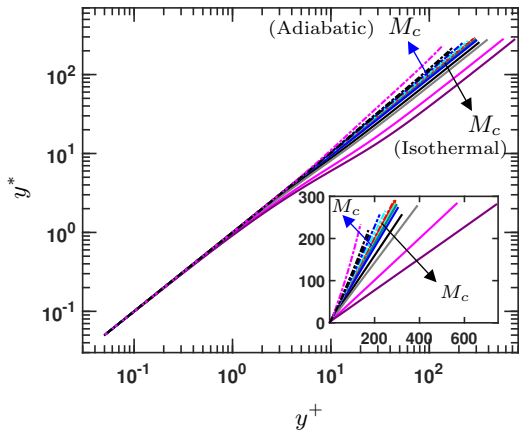

Because of the different scaling observed with wall and semilocal units,
a natural question is, thus, on the relation between these two 
normalized distances to the wall.
In \rfig{semilocal}, we show 
the wall-normal coordinate in semi-local units ($\yst$) against 
the wall-normal coordinate in viscous units ($\yp$). 
The two normalizations
are virtually the same in the viscous sublayer for 
the given range of Mach numbers.
Further away from the wall, 
isothermal and adiabatic walls lead to opposite trends
 when the Mach number is increased.
These can be explained as follows.
From \rfig{thmean} (a) and (c), we found that away 
from the wall, viscosity increases and density decreases   
as the Mach number is increased for isothermal cases while 
opposite trends are observed for adiabatic cases. Thus, the 
ratio $\sqrt{\ol{\rho}}/\ol{\mu}$ decreases with increasing Mach number for
isothermal
cases, while this ratio increases for adiabatic cases when $M_c$ is increased. 
Thus, following this trend,
$\yst$ decreases with $M_c$ for fixed $\yp$ 
for isothermal cases while 
it increases with $M_c$ for adiabatic cases as shown in
\rfig{semilocal}(a). We can also see that $\yst$ at the centerline ($\yst_c$) 
reaches a range of values of
$\yst_c\approx 220-290$ for all Mach numbers and WTBCs.
This range is much wider for 
wall units, $\yp_c \approx 140-750$, which seems to support the idea that
semi-local units provide a better
self-similar normalization than wall units.
However, we note that this may be, in part, due to the fact that 
simulations were conducted with an approximately constant $Re_{\tau}^*$
\citep{TL2016}.
Further simulations at a wide range of $Re_{\tau}^*$ are needed to 
provide a more definite assessment of this claim.

\section{Effects of thermal boundary conditions on turbulent stresses}
\label{sec:stresses}
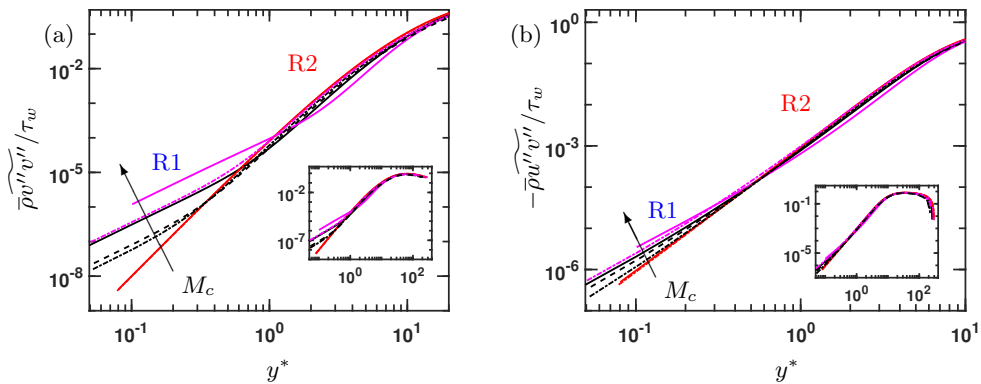
\begin{figure}
\centering
\includegraphics[width=.48\textwidth]{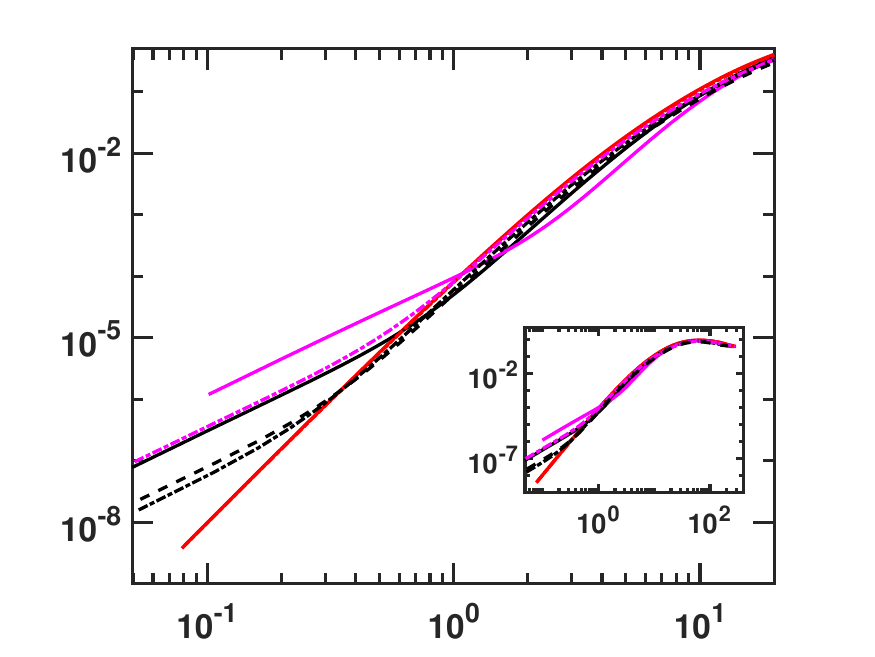}
\begin{picture}(0,0)
      \put(-180,116){ (a)}
   \put(-190,50){\rotatebox{90}{ {$\ol{\rho} \wt{v''v''}/\tau_w$}}}
 \put(-96, -9.5){ {$y^*$}}
   \put(-128,30){\vector(-1,2){20}}
   \put(-128,22){ $M_c$}
 \put(-136,68){\colb{R1}}
 \put(-85,105){\colr{R2}}
\end{picture}
\includegraphics[width=.48\textwidth]{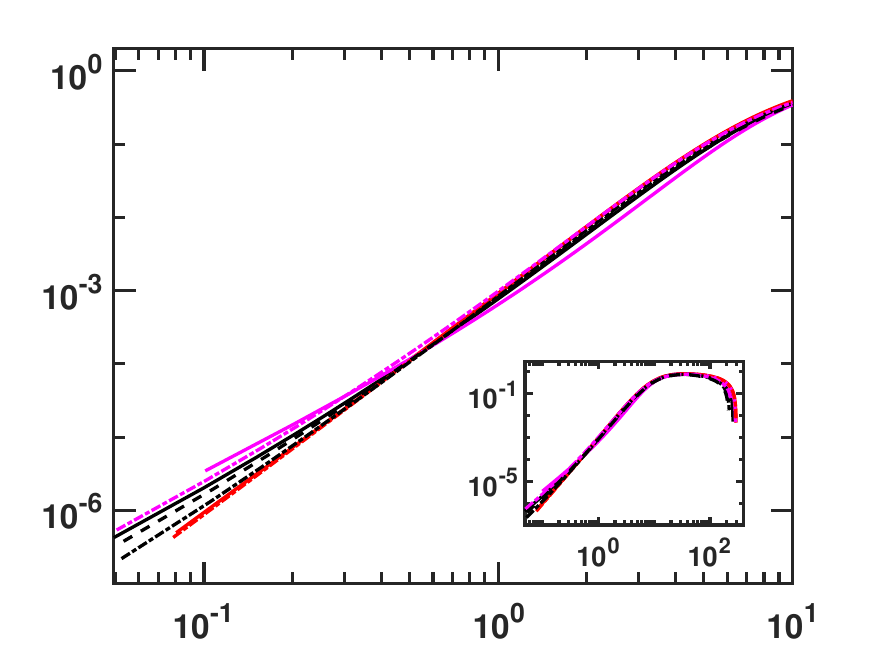}
\begin{picture}(0,0)
      \put(-195,116){ (b)}
   \put(-190,50){\rotatebox{90}{ {$-\ol{\rho} \wt{u''v''}/\tau_w$}}}
 \put(-92, -9.5){{$y^*$}}
   \put(-137,28){\vector(-1,2){12}}
   \put(-137,20){ $M_c$}
 \put(-140,50){\colb{R1}}
 \put(-90,90){\colr{R2}}
\end{picture}
\vskip 0.4 cm
	\caption{(a)-(b) Density-scaled Reynolds stresses distributions
	versus semi-local wall-normal coordinate for isothermal 
        (---), adiabatic (-\,$\cdot$\,-) and pseudo-adiabatic (-\,-\,-) cases. 
	Red, black and magenta correspond to $M_c \approx 0.23$, 
$M_c \approx 1.2$ and $M_c \approx 1.9$ respectively. Insets show same profiles up to $y^*\approx 300$.}
\label{fig:rij}
\end{figure}

The wall-normal coordinate in semi-local units, $\yst$
along with local density-weighted averaging
have been widely used to try to collapse turbulent stresses in compressible
wall-bounded flow with varying WTBC, with their incompressible counterparts
\citep{HCB1995,FSF2004,MTN2004,TL2016,MP2016,ZDC2018}.
We have recently shown \cite{BDB2022}, however, that semi-local scaling 
is not able to collapse turbulent stresses
$R_{\alpha\beta}^* \equiv \ol{\rho} \wt{\alpha''\beta''}/\tau_w$
($\alpha$ and $\beta$ are velocity components;
e.g., $R_{uv}^* \equiv \ol{\rho} \wt{u''v''}/\tau_w$)
or the wall-normal turbulent heat flux, 
$ R_{vT}^* = \ol{\rho} \wt{v''T''}/(\rho_wu_\stau T_\stau)$ 
close to an isothermal wall in turbulent channels
for centerline Mach numbers
ranging from the
incompressible limit to supersonic regimes.
This can also be observed here in, e.g., \rfig{rij}
(a)(b). 

In \rfig{rij} (a) and (b), we show $R_{vv}^*$ and $R_{uv}^*$ respectively
for three Mach numbers, 
$M_c \approx 0.23, 1.2$ and $1.9$, for 
both isothermal (solid lines) and adiabatic (dash-dotted lines) walls. The figure 
also include one pseudo-adiabatic case (dashed line) at $M_c \approx 1.2$. 
At the lowest Mach number ($M_c \approx 0.23$), 
turbulent stresses ($R_{vv}^*$, $R_{uv}^*$) collapse well 
for isothermal and adiabatic walls
suggesting no appreciable WTBC effect as one approaches the incompressible limit.
As the Mach number is increased, however, we can clearly observe 
differences between isothermal, adiabatic and pseudo-adiabatic cases 
for $R_{vv}^*$ and $R_{uv}^*$ which are apparent for $M_c \approx 1.2$ and beyond.
This effect is especially strong in the viscous sub-layer where we can 
clearly see higher normal Reynolds stresses close to
isothermal (solid line) than to adiabatic (dashed-dotted line) walls.
However, one can also observe that
some Mach number effects are similar in isothermal cases 
and adiabatic cases. 
Investigating these differences and similarities 
are the main focus of the current work.

Three observations can be made. 
First, in the region adjacent to the wall, indicated by R1
in \rfig{rij},
we can see power-law behavior for both 
$R_{uv}^*$ and $R_{vv}^*$ with exponents that
decrease with $M_c$ for adiabatic cases and, as observed before, 
isothermal cases (\cite{BDB2022}).
The slope of $R_{uv}^*$ in R1, however, does change with WTBC when
$M_c$ is kept constant.
This WTBC effect is much weaker for $R_{vv}^*$.
Second, $R_{uv}^*$ and $R_{vv}^*$ transition to another scaling regime,
indicated as R2 in \rfig{rij}, with much
weaker WTBC and $M_c$ effects. Finally, the transition location changes with both
$M_c$ and WTBC. 
Taken together, these general observations suggest 
that significant WTBC and Mach number effects are observed close the wall
as Mach number increases.

\begin{table}
\begin{center}
\def~{\hphantom{0}}
\begin{tabular}{|l|c|c|c|}
\hline
      &  $\ol{v'v'}$ & $\ol{u'v'}$ & $\ol{v'T'}$  \\ 
      &&& isothermal $\hspace{0.2cm}$  adiabatic $\hspace{0.2cm}$ pseudo-adiabatic\\
	\hline
     solenoidal &  4 & 3 & 3 $\hspace{1.5cm}$ 2 $\hspace{1.5cm}$ 3\\
     non-solenoidal   &  2 & 2 & 2 $\hspace{1.5cm}$ 1 $\hspace{1.5cm}$ 2\\
     \hline
\end{tabular}
\caption{
Exponents $\mygam{}{\alpha\beta}$ for near wall asymptotic behavior
for $R_{\alpha\beta}$ ($\alpha$ and $\beta$ are $u$, $v$ or $T$).
}
\label{tab:gamma}
\end{center}
\end{table}

\begin{table}
\begin{center}
\def~{\hphantom{0}}
\begin{tabular}{|l|c|c|c|}
\hline
          &  Isothermal & Adiabatic & Pseudo-adiabatic  \\ 
      	     & R1$\hspace{0.2cm}$R2& R1$\hspace{0.2cm}$R2& R1$\hspace{0.2cm}$R2    \\ \hline
	wall & 
\tikz \draw[blue,fill=white] (1.5,0.75)--(1.5,0.9)--(1.65,0.9)--(1.65,0.75)--cycle;
$\hspace{0.2cm}$
\tikz \draw[red,fill=white] (1.5,0.75)--(1.5,0.9)--(1.65,0.9)--(1.65,0.75)--cycle;
& 
\tikz \draw[blue,fill=blue] (1.5,0.75)--(1.5,0.9)--(1.65,0.9)--(1.65,0.75)--cycle;
$\hspace{0.2cm}$
\tikz \draw[red,fill=red] (1.5,0.75)--(1.5,0.9)--(1.65,0.9)--(1.65,0.75)--cycle;
  & 
\tikz \draw[royalblue,fill=royalblue] (1.5,0.75)--(1.5,0.9)--(1.65,0.9)--(1.65,0.75)--cycle;
$\hspace{0.2cm}$
\tikz \draw[salmon,fill=salmon] (1.5,0.75)--(1.5,0.9)--(1.65,0.9)--(1.65,0.75)--cycle;
   \\
      semi-local   &
\tikz \draw[blue,fill=white] (1.5,0.75)--(1.6,0.9)--(1.7,0.75)--cycle;
$\hspace{0.2cm}$
\tikz \draw[red,fill=white] (1.5,0.75)--(1.6,0.9)--(1.7,0.75)--cycle;
& 
\tikz \draw[blue,fill=blue] (1.5,0.75)--(1.6,0.9)--(1.7,0.75)--cycle;
$\hspace{0.2cm}$
\tikz \draw[red,fill=red] (1.5,0.75)--(1.6,0.9)--(1.7,0.75)--cycle;
&
\tikz \draw[royalblue,fill=royalblue] (1.5,0.75)--(1.6,0.9)--(1.7,0.75)--cycle;
$\hspace{0.2cm}$
\tikz \draw[salmon,fill=salmon] (1.5,0.75)--(1.6,0.9)--(1.7,0.75)--cycle;
   \\
   \hline
\end{tabular}
\caption{
Marker styles used for exponents $\gamvv$ and $\gamuv$ for different 
WTBCs and scaling regimes.
}
\label{tab:marker}
\end{center}
\end{table}

The near-wall asymptotic behavior of turbulent stresses 
can be theoretically estimated by expanding the constituent velocity
components as Taylor series expansions in $y$:
\be
  u'=a_u + b_u y + c_u y^2 +\dots, \quad
  v'=a_v + b_v y + c_v y^2 +\dots
\ee
The coefficients $a_\alpha$ for $\alpha=u$ and $v$
are identically zero
due to the no-slip boundary condition 
at the wall.
The other coefficients are given by
$b_v=\partial v'/\partial y$, and $c_v=(1/2)\partial^2 v'/\partial y^2$,
and similarly for $u$.
If the flow is incompressible (solenoidal), mass conservation combined 
with the no-slip condition at the wall leads to an additional
constraint in the wall-normal velocity component,
namely, $\partial v'/\partial y=b_v=0$.
On the other hand, if the flow is non-solenoidal, $b_v \ne 0$. 
By taking the product between the expansions of different components
and averaging, one can formulate Reynolds averaged turbulent stresses 
($R_{\alpha\beta} \equiv \ol{\alpha'\beta'}/u_{\tau}^2$),
resulting in near-wall scaling laws of the form
$R_{\alpha\beta} \approx \sigma_{\alpha\beta}y^{\gamma_{\alpha\beta}}$
with exponents summarized in \rtab{gamma}. 
These theoretical exponents are the same for $R_{\alpha\beta}$
and $R_{\alpha\beta}^*$ given that density has a finite value at the wall.
From \rtab{gamma}, we see that the solenoidal and non-solenoidal
exponents are different for turbulent stresses containing a
wall-normal velocity component. 
As in \cite{BDB2022}, we investigate 
exponents ($\gamma_{\alpha\beta}$) and pre-factors ($\sigma_{\alpha\beta}$)
but extending the analysis to include WTBC effects.

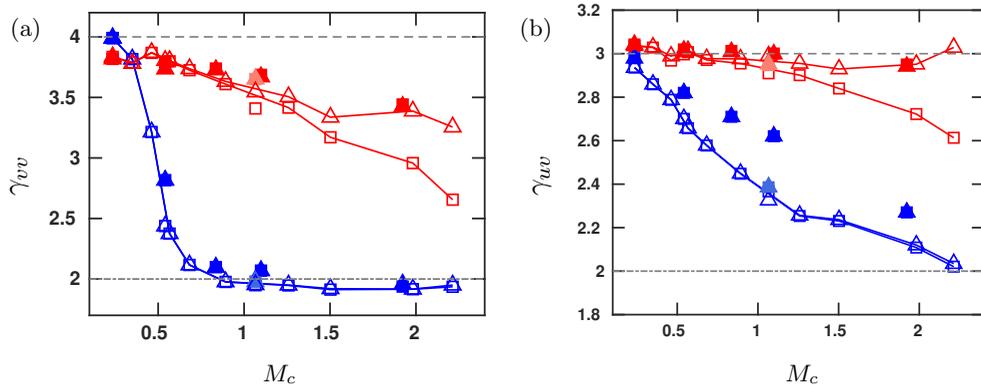
\begin{figure} 
\centering
\includegraphics[width=.50\textwidth]{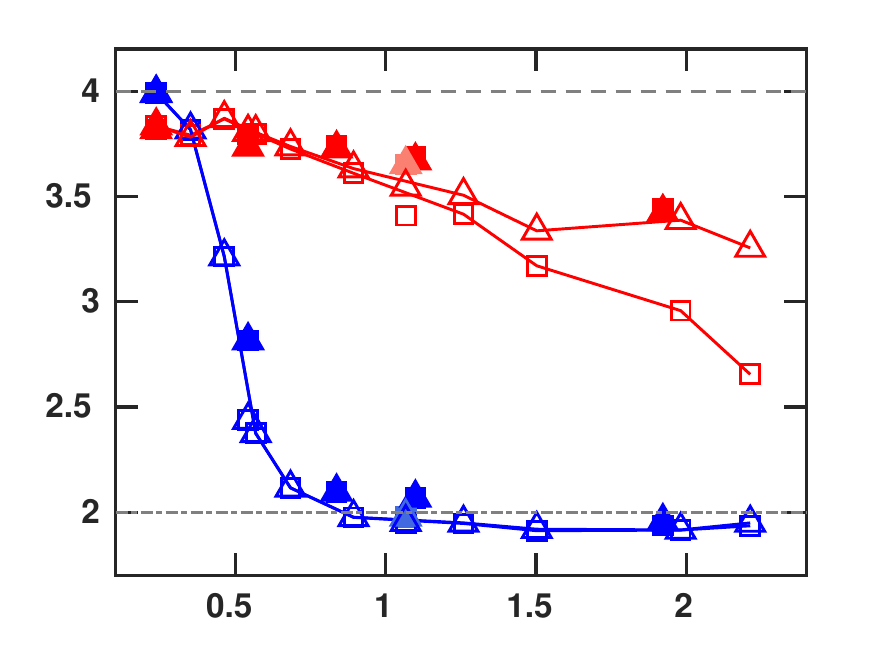}
\begin{picture}(0,0)
    \put(-200, 120){(a)}
    \put(-200,60){\rotatebox{90}{\large {$\gamma_{vv}$}}}
    \put(-105, -9.5){{$M_c$}}
\end{picture}
\includegraphics[width=.47\textwidth]{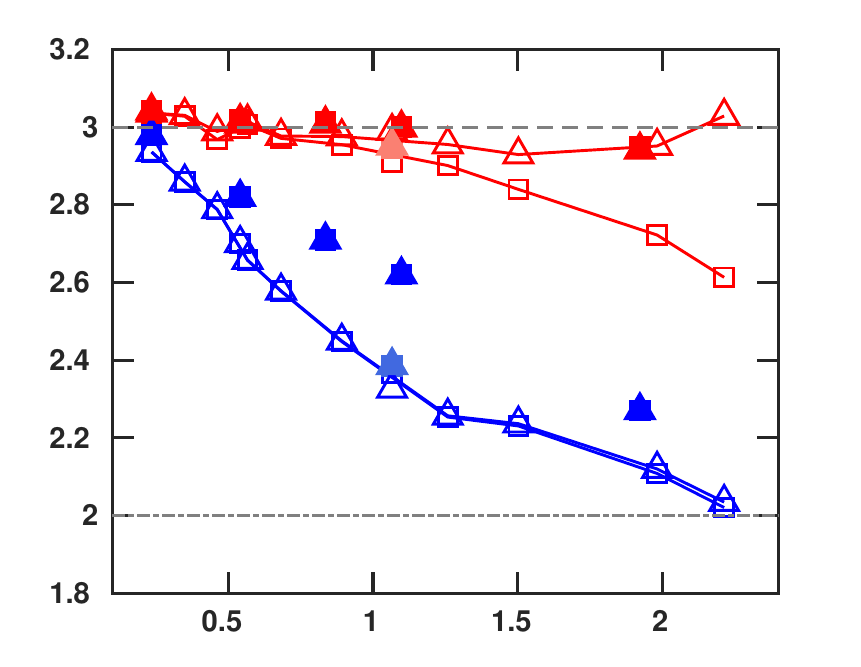}
\begin{picture}(0,0)
    \put(-195, 120){(b)}
    \put(-190,60){\rotatebox{90}{\large {$\gamma_{uv}$}}}
    \put(-95,-9.5){{$M_c$}}
\end{picture}
\vskip 0.15 cm
	\caption{ Power law exponents for (a) wall-normal Reynolds stress 
	(b) Reynolds shear stress plotted against centerline Mach number. 
	Horizontal gray lines for 
	solenoidal (-\,-\,-) and non-solenoidal (-\,$\cdot$\,-)
	asymptotic exponents (\rtab{gamma}).
	Markers in all panels (\rtab{marker}): $\square$ indicates wall normalizations ($R_{\alpha\beta} =
	\sigma_{\alpha\beta}^+ (y^+)^{\gamma_{\alpha\beta}^+}$), 
	$\triangle$ indicates semi-local normalizations ($R_{\alpha\beta}^* =
	\sigma_{\alpha\beta}^* (y^*)^{\gamma_{\alpha\beta}^*}$) for isothermal 
(empty markers), pseudo-adiabatic (light-filled markers) and adiabatic 
(dark-filled markers) cases.
	Blue and red markers correspond to R1 and R2 regions, respectively.
	The solid line in all panels connects isothermal data for comparison.}
\label{fig:gamvv}
\end{figure}

Following \cite{BDB2022}, 
we fit power laws in 
regions R1 and R2, as shown in \rfig{rij} for both wall ($R_{\alpha\beta}$ 
versus $y^+$: $\square$) and semi-local
($R_{\alpha\beta}^*$ versus $y^*$: $\triangle$) 
normalizations, to obtain 
($\gamma^+_{\alpha\beta}$, $\sigma^+_{\alpha\beta}$)  
and ($\gamma^*_{\alpha\beta}$, $\sigma^*_{\alpha\beta}$)
respectively for all cases in our database. 
In \rfig{gamvv} (a), we show the exponent
$\gamma_{vv}$
for isothermal (empty markers), adiabatic (dark-filled markers) 
and pseudo-adiabatic (light-filled markers)
wall conditions as a function of $M_c$.
The theoretical asymptotic values in \rtab{gamma} are expected to be 
attained for exponents in R1 (blue symbols) which are the closest to the wall.
On changing thermal wall conditions, the difference between 
$\gamma_{vv}$ in R1
is small for the same centerline Mach number 
except for $M_c=0.5$ 
where the adiabatic case has a slightly larger exponent. 
The exponent $\gamma_{vv}$ approaches its solenoidal and non-solenoidal
limiting behavior
(see \rtab{gamma}) for $M_c \lesssim 0.2$ and $M_c \gtrsim 0.8$, respectively.
Between these two limits there is a smooth transition with $M_c$
for both isothermal and adiabatic cases. 

In \rfig{gamvv}(b) we show the exponents for the shear Reynolds 
stress,
$\gamma_{uv}$, versus $M_c$
and observe a much stronger influence of thermal boundary conditions 
with larger values of $\gamma_{uv}$ in R1 for adiabatic 
cases at all Mach numbers.
The pseudo-adiabatic case appears to match the isothermal case, which may not
be completely unexpected given that in this case
we also impose a constant 
temperature at the wall. 
This may indicate that $\gamma_{uv}$ is independent of $T_w$ since 
exponents for isothermal and pseudo-adiabatic are very close to each
other even though wall temperature is markedly different.
Furthermore, this may also suggest that differences in exponents for 
isothermal and adiabatic cases are not due to differences in wall temperature.
The values obtained for the R1 exponents, however, are independent
of whether one uses wall or semilocal units for all WTBCs.
This is in line with the theoretical behavior discussed earlier.

In R2, semi-local normalization provides a better collapse 
of exponents with different WTBCs.
This can be seen in \rfig{gamvv}(a) where we 
see that, for a fixed $M_c$,
there are negligible differences between $\gamvvs$ for 
isothermal (red empty triangles), 
adiabatic (red filled triangles), and 
pseudo-adiabatic (light red filled triangles) cases. 
Note also that
$\gamvvp$ and $\gamvvs$ in R2 are the same
for all Mach numbers for adiabatic cases 
but not for isothermal cases. 
For isothermal cases,
when $M_c$ is roughly above unity, 
$\gamvvp$ and $\gamvvs$
differ. This can be understood by noting
that the temperature and density gradients are higher near the isothermal wall
than the adiabatic wall. Therefore, in adiabatic cases,
local density and viscosity are closer to wall values as compared to those in
isothermal cases (also seen in \rfig{thmean}(a)(c)).  
Similar behavior is observed for $\gamuv$.
In \rfig{rij}, we found that turbulent stresses in R2 are less affected 
by variations in Mach number when semi-local normalizations are used for
different WTBCs.
This is consistent with 
the results in \rfig{gamvv}(a)(b), where we see a very weak 
$M_c$ effect on $\gamuvs$ (and to a lesser degree on $\gamvvs$)
 for all WTBCs.
In general, though, we observe a weaker $M_c$ dependence for adiabatic
than isothermal walls for exponents in wall units.

In addition to obtaining exponents for isothermal cases from simulations 
where both walls are isothermal and at the same temperature, we also obtain the 
exponents close to the isothermal wall from simulations with different thermal 
boundary conditions (pseudo-adiabatic or adiabatic) on the other wall. 
The exponents $\gamma_{vv}$ and $\gamma_{uv}$
in R1 near the isothermal wall was found to, in fact, be
independent of the boundary condition of the other wall, 
indicating that the near-wall 
asymptotic behavior is not significantly affected by the WTBC 
on the non-identical wall. 

\begin{figure} 
\centering
\includegraphics[width=.48\textwidth]{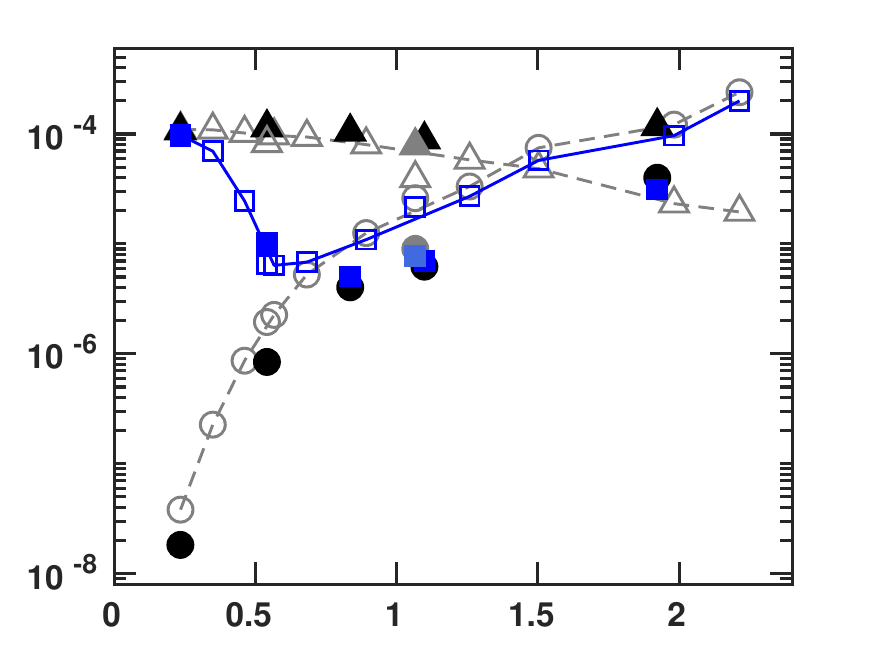}
\begin{picture}(0,0)
    \put(-190, 120){(a)}
    \put(-195,64){\rotatebox{90}{\large {$\sigma_{vv}$}}}
    \put(-98, -9.5){\small {$M_c$}}
\end{picture}
\includegraphics[width=.48\textwidth]{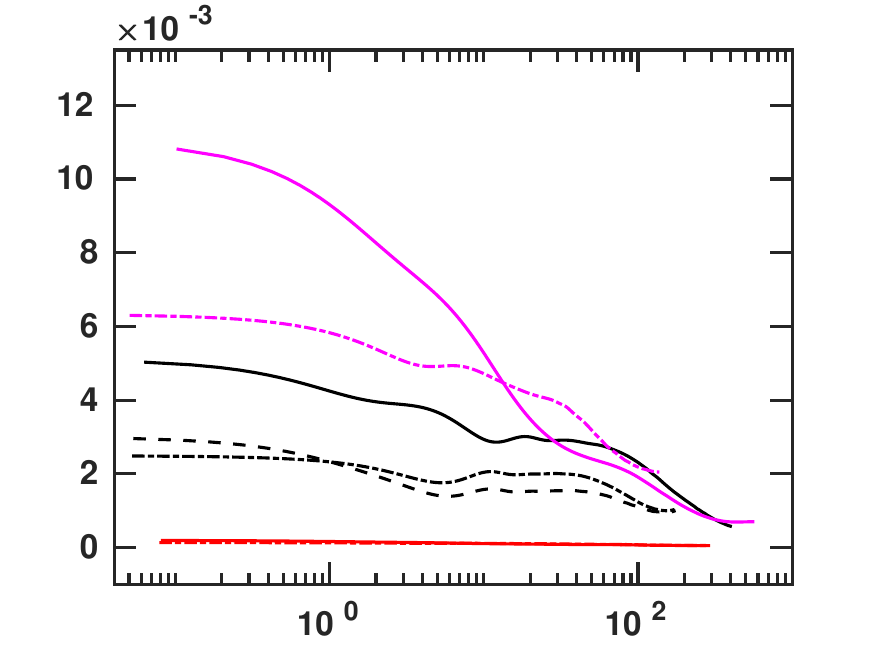}
\begin{picture}(0,0)
    \put(-192, 120){\small (b)}
    \put(-195,64){\rotatebox{90}{\large {$\theta_{rms}^+$}}}
     \put(-140,18){\vector(0,2){100}}
   \put(-136,110){  $M_c$}
    \put(-98, -9.5){ {$y^+$}}
\end{picture}
\vspace {4 mm}
 \caption{(a) Prefactor $\sigma_{vv}$ (squares) in R1 and normalized 
	coefficients in Taylor expansion for $R_{vv}$,   non-solenoidal,
	$\ol{b_v^2}\nu_w^2/u_\tau^4$ (circles) and solenoidal,
	$\ol{c_v^2}\nu_w^4/4 u_\tau^6$ (triangles) 
	against $M_c$ for different WTBCs. Markers are given in \rtab{sigmar}.
	(b) Distribution of root-mean-squared dilatation 
	with wall-normal coordinate for isothermal 
        (---), adiabatic (-\,$\cdot$\,-) and pseudo-adiabatic (-\,-\,-) cases.
	Red, black and magenta correspond to $M_c \approx 0.23$, 
	$M_c \approx 1.2$ and $M_c \approx 1.9$ respectively.  }
\label{fig:dila}
\end{figure}
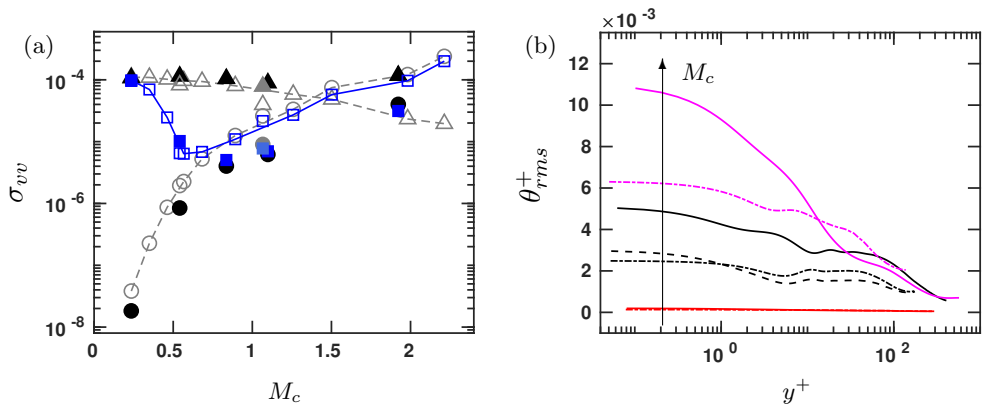

\begin{table}
\begin{center}
\def~{\hphantom{0}}
\begin{tabular}{|l|c|c|c|}
\hline
           &  Isothermal & Adiabatic & Pseudo-adiabatic  \\ 
        \hline
      	   $\sigma_{vv}$ & 
          \tikz \draw[blue,fill=white] (1.5,0.75)--(1.5,0.9)--(1.65,0.9)--(1.65,0.75)--cycle;
&
\tikz \draw[blue,fill=blue] (1.5,0.75)--(1.5,0.9)--(1.65,0.9)--(1.65,0.75)--cycle;
& 
\tikz \draw[royalblue,fill=royalblue] (1.5,0.75)--(1.5,0.9)--(1.65,0.9)--(1.65,0.75)--cycle;
   \\
          $\ol{b_v^2}\nu_w^2/u_\tau^4$ &
           \tikz \filldraw[color=gray, fill=white] (0,0) circle (2.5pt);
           &
           \tikz \filldraw [black] (0,0) circle (2.5pt);
           &
          \tikz \filldraw [gray] (0,0) circle (2.5pt);
          \\
          $\ol{c_v^2}\nu_w^4/4 u_\tau^6$ &     
          \tikz \draw[gray,fill=white] (1.5,0.75)--(1.6,0.9)--(1.7,0.75)--cycle;
&
\tikz \draw[black,fill=black] (1.5,0.75)--(1.6,0.9)--(1.7,0.75)--cycle;
& 
\tikz \draw[gray,fill=gray] (1.5,0.75)--(1.6,0.9)--(1.7,0.75)--cycle;
   \\
 \hline
\end{tabular}
\caption{
Marker styles used for prefactors and coefficients in Taylor series expansion for $R_{vv}$ for different 
WTBCs.
}
\label{tab:sigmar}
\end{center}
\end{table}

Written out explicitly, the first three terms in the expansion of the 
wall-normal Reynolds stress is 
$\ol{v'v'}=\ol{b_v^2}y^2 + (\ol{b_vc_v}/2)y^3+(\ol{c_v^2}/4+\ol{b_vd_v}/6)y^4+{\cal O}(y^5)$.
As discussed above, when the flow is incompressible $b_v=0$ and the $y^4$
term
dominates; when the flow is compressible, one expects the $y^2$
term to dominate. However, this would also depend on the prefactors 
involved, $\ol{b_v^2}$ and $\ol{c_v^2}$, which in a particular 
region may make one term dominates over the other. 
In \rfig{dila}(a) we show these prefactors normalized with wall units,
that is,
$\ol{b_v^2}\nu_w^2/u_\tau^4=
\ol{(\partial v'/\partial y)_w^2}\nu_w^2/u_\tau^4$ (circles)
and 
$\ol{c_v^2}\nu_w^4/4 u_\tau^6=
\ol{(\partial^2 v'/\partial y^2)_w^2}\nu_w^4/4 u_\tau^6$ (triangles)
computed using the derivatives from DNS data at the wall.
In the same plot, we also include the prefactor $\sigma^+_{vv}$
obtained from the fits $R_{vv}\ap \sigma^+_{vv}(y^+)^\gamvvp$
as described above.
We can clearly see that $\sigma^+_{vv}$ (squares) tends to the 
solenoidal (triangles) and non-solenoidal (circles) analytical values for 
$M_c\lesssim 0.2$ and $M_c\gtrsim 0.8$, respectively
for isothermal (empty markers) and adiabatic (dark-filled markers) 
wall conditions. Pseudo-adiabatic (light-filled marker) 
case with $M_c \approx 1.2 $ also follows analytical
non-solenoidal value.
These observations are consistent with the behavior of exponents obtained from
the fit. 
The value of $\sigma^+_{vv}$ (squares) is also found to be lower 
for adiabatic (dark-filled) than isothermal (empty) cases at $M_c \gtrsim 0.8$.
We finally note that, at high Mach numbers, the dominant prefactor is 
the one involving $b_v$ which for no-slip walls, 
is equal to the level of dilatation motions at the wall
\citep{BDB2022}. Thus, from a purely kinematic standpoint, 
the particular scaling laws observed will
depend only on dilatation (i.e.~$b_v$) regardless of how those
dilatations are generated.

It is known that
different levels of dilatation at the wall can be generated either by
changing the centerline Mach number \citep{BDB2022} or 
thermal boundary condition at the wall (\cite{XWYLC2021}).
This is also clear in \rfig{dila}(b), where we observe that the 
level of dilatational motions at the wall
is different for different Mach numbers and WTBCs. Dilatation levels are weaker
 for adiabatic than isothermal walls with the same $M_c$. 
Pseudo-adiabatic walls have intermediate dilatation levels
close to the wall. 
As previously stated, dilatation is a key factor governing the scaling laws,
and one may, thus, expect better collapse of different 
statistics when using the dilatational content as a normalizing 
parameter.
This general concept of universality based on the level of dilatational motions
independent of the specific mechanism that 
generated them was indeed recently proposed \citep{DJ2020}
though only for homogeneous flows.

\begin{figure} 
\centering
\includegraphics[width=.48\textwidth]{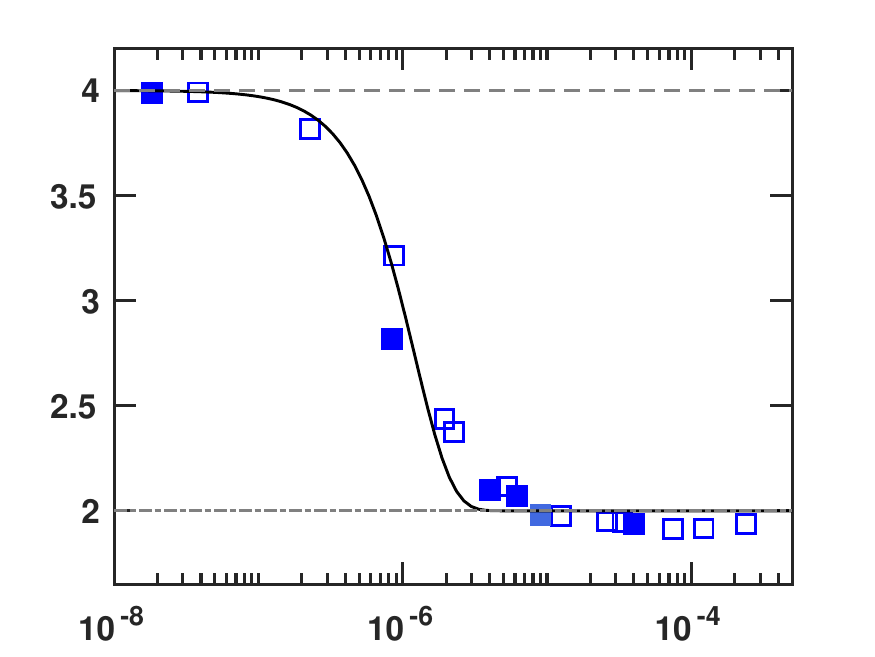}
\begin{picture}(0,0)
\put(-185, 120){(a)}
    \put(-190,62){\rotatebox{90}{\large {$\gamma_{vv}$}}}
    \put(-105, -9.5){{$\dilwrms^+$}}
\end{picture}
\includegraphics[width=.48\textwidth]{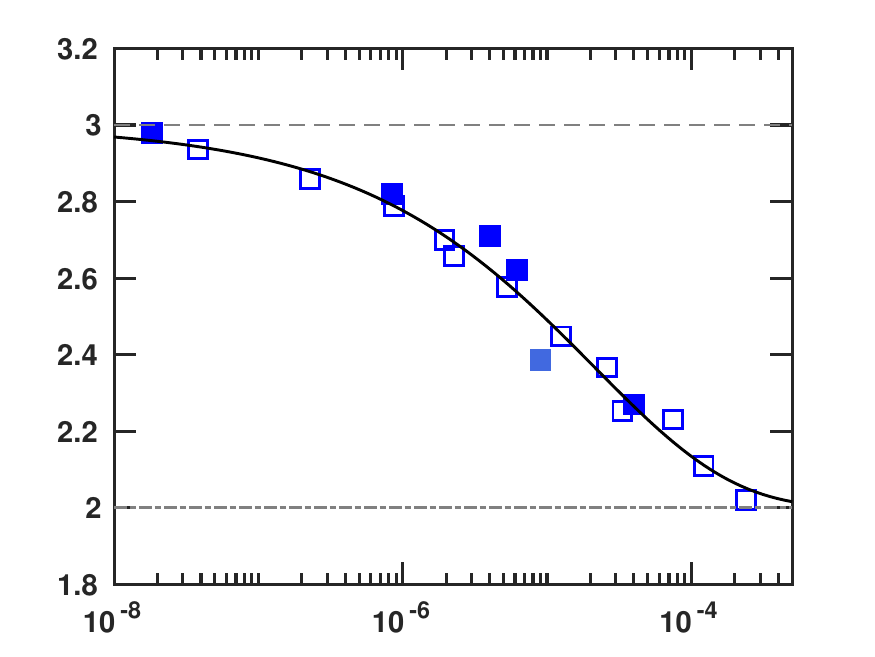}
\begin{picture}(0,0)
    \put(-190, 120){(b)}
    \put(-190,62){\rotatebox{90}{\large {$\gamma_{uv}$}}}
    \put(-105, -9.5){{$\dilwrms^+$}}
\end{picture}
\vspace{6 mm}
	\caption{Power-law exponents in R1 for (a) wall-normal turbulent stress 
	(b)turbulent shear stress plotted against r.m.s of dilatation 
 	at the wall. Markers as in \rtab{marker}.
Horizontal gray lines for solenoidal (- - -)
and non-solenoidal (- · -) asymptotic exponents (table 2). 
Solid lines are scalings, 
 	(a) $2+2\exp(-10^{10}{\dilwrms^+}^{1.69})$ 
 	(b) $2+\exp(-126{\dilwrms^+}^{0.45})$.    
 }
\label{fig:fgvtheta}
\end{figure}
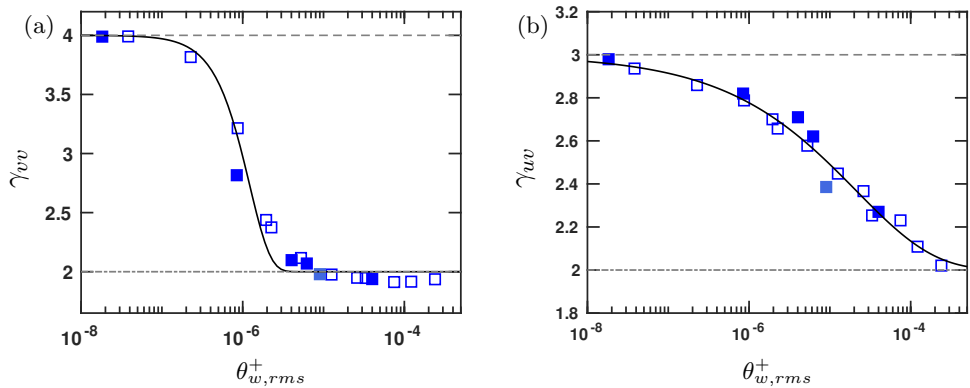

To test these concepts, in \rfig{fgvtheta} (a)(b) we show 
the exponents as a function of the r.m.s.\
of dilatation at the wall normalized
with wall units, $\dilwrms^+$.
We clearly see a better collapse 
of exponents than in the corresponding panels (a) and (b) of \rfig{gamvv},
supporting the idea that dilatational levels,
regardless of how they are generated, provide the  
appropriate scaling parameter for near-wall behavior 
at high speeds.
This is consistent with \cite{DJ2020} where the use of
dilatational content as a governing parameter yielded  
a universal behavior for 
a number of statistics including pressure variance, dissipation,
and skewness of the velocity gradients.
From a modeling perspective, it may be useful to parametrize 
these
seemingly universal curves. 
We have found that these curves can be represented reasonably well 
with simple exponentials in $\dilwrms^+$, which are included in 
\rfig{fgvtheta}(a)(b) and noted in its caption.

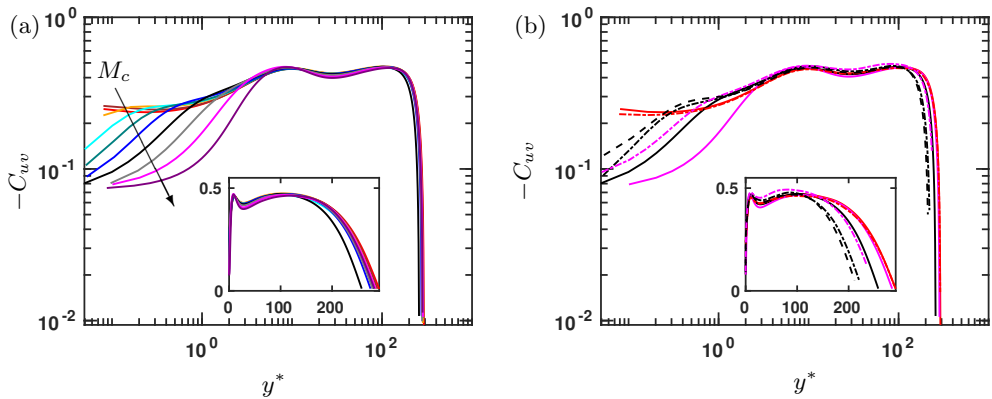
\begin{figure} 
\centering
\includegraphics[width=.49\textwidth]{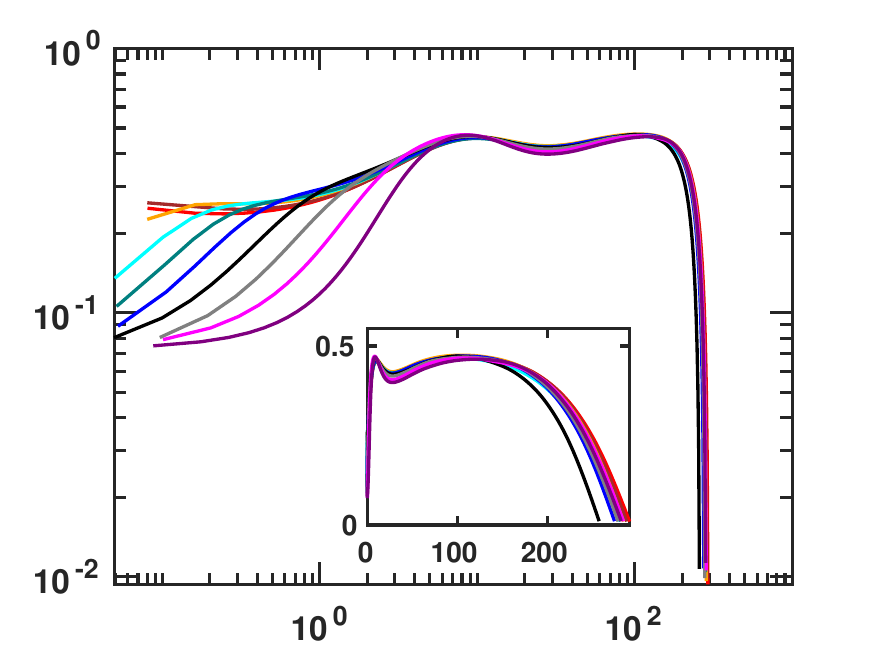}
\begin{picture}(0,0)
    \put(-195, 125){(a)}
    \put(-195,58){\rotatebox{90}{{$-C_{uv}$}}}
    \put(-100, -9.5){{$y^*$}}
   \put(-155,104){\vector(1,-2){22}}
   \put(-162,108){$M_c$}
\end{picture}
\includegraphics[width=.49\textwidth]{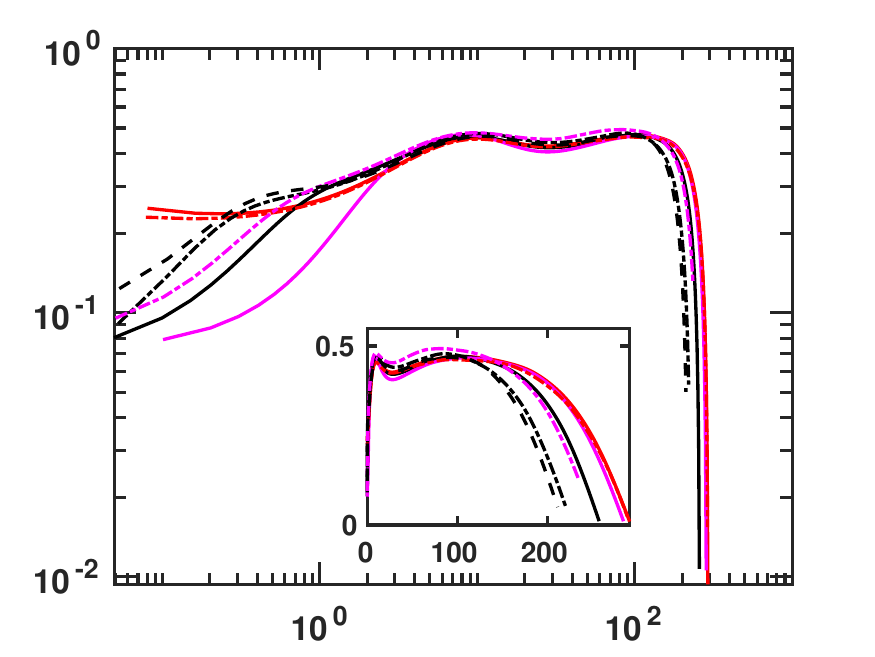}
\begin{picture}(0,0)
    \put(-5, 135){(b)}
    \put(-5,72){\rotatebox{90}{{$-C_{uv}$}}}
    \put(100, 1.5){{$y^*$}}
    \vspace{6mm}
\end{picture}
	\caption{
	Correlation coefficient for $R_{uv}$ (a) Isothermal wall
        (b) Isothermal (---), adiabatic  (-\,$\cdot$\,-) and pseudo-adiabatic 
(-\,-\,-) walls.  Inset contains the same data in linear scales.  Colors as in \rtab{dns1}. }
\label{fig:cuv}
\end{figure}

On comparing \rfig{gamvv}(a) with (b), 
we find that the transition from the low to the high Mach number limit in R1
for $\gamma_{uv}$ is smoother than that of $\gamma_{vv}$ for isothermal as well
as adiabatic cases (adiabtic cases exhibit an even slower transition than isothermal
cases).
A similar observation can also be made from
\rfig{fgvtheta} where the transition (with levels of dilatation at the wall
in this case)
is smoother for $\gamuv$ as compared to $\gamvv$.
This suggests a slow decorrelation between $u'$ and $v'$ as compressibility
levels increase close to the wall.
To study this, we show in \rfig{cuv}(a) the correlation coefficient 
$C_{uv} \equiv \ol{u'v'}/u_{rms}v_{rms}$ 
for all isothermal cases in the database. 
We similarly define the
correlation coefficient $C_{\alpha\beta}$ for arbitrary
variables $\alpha$ and $\beta$ as
\be
C_{\alpha\beta} \equiv \frac{\ol{\alpha'\beta'}}{\alpha_{rms}\beta_{rms}} 
\label{eq:calbt}
\ee
We see that for the lowest Mach numbers,
$C_{uv}$ is relatively constant close to the wall 
($y^*\lesssim 1$).
As $M_c$ increases, the overall magnitude of the correlation is 
reduced in this region, though all the lines seem to approach,
a region of relatively constant correlation of about 
0.45, a value consistent with those observed 
in supersonic boundary layers (\cite{SHH2015}).
The distance from the wall at which this region starts, however,
increases with $M_c$, indicating that 
compressibility effects are felt at increasing distance from the wall
as the Mach number increases.
The increasing decorrelation close to the wall with $M_c$ 
has also been observed in \cite{SCG2017}, an effect that 
was also found to be independent of Reynolds number. 
This near-wall decorrelation that becomes stronger as $M_c$
increases suggests that while a simple product of Taylor expansions 
can describe diagonal stresses (e.g.~$R_{uu}$ or $R_{vv}$), 
this is not the case for off-diagonal stresses ($R_{uv}$) which comprise the 
correlation between two different variables.
In particular, we see that for low and high $M_c$, the correlation 
$C_{uv}$ is relatively constant close to the wall, though at different 
levels. It is at intermediate Mach numbers that $C_{uv}$ presents 
a positive slope in this region.
Thus, because
$\ol{u'v'}=C_{uv}\urms\vrms$ 
we can see how the R1 exponent for $R_{uv}$, would be close to the 
sum of the exponents for $\urms$ and $\vrms$ for low and high $M_c$
while it would be larger 
at intermediate $M_c$. This explains, then, why 
the transition from the solenoidal to the non-solenoidal asymptotes 
is smoother for $R_{uv}$ than for the case of diagonal stresses.
At the  centerline of the channel, $C_{uv}$ vanishes due
to reflective symmetry across the centerline plane, which is seen as a rapid
decrease in the correlation in the figure at high values of $y^*$.

To assess the effect of WTBC,
in \rfig{cuv}(b) we show the correlation coefficient for different boundary
conditions and three Mach numbers, $M_c \approx 0.23$, $1.2$, and $1.9$.
As before, we see that $C_{uv}$ is relatively flat at the lowest $M_c\ap 0.23$
and for distances below $y^*\sim O(1)$,
with very little WTBC effect. The same weak dependence on WTBC
is observed at $y^*$ beyond, say, 4, where $C_{uv}$ approaches the constant value
discussed above.
As the Mach number is increased, however, there are observable differences
between isothermal, adiabatic and pseudo-adiabatic walls.
In particular, we see that 
isothermal walls (black solid line) create a stronger 
decorrelation between $u$ and $v$ than adiabatic walls (black dashed-dotted
line) for $M_c\gtrsim 1.2$ and pseudo-adiabatic (black dashed line) for $M_c \ap 1.2$.
In addition, there are differences in the slope for $C_{uv}$ close to the wall 
between adiabatic and isothermal cases, especially for $M_c \approx 1.2$ 
which also seem to contribute to the difference in power-law behavior for 
these two WTBCs. This is clearly evident in \rfig{gamvv}(b), 
where $\gamvv$ for $M_c \approx 1.2$ seems to have the largest difference 
between isothermal and adiabatic cases. Moreover, the distance from the wall
at which constant region of $C_{uv}$ starts, is larger for isothermal than
adiabatic cases. 

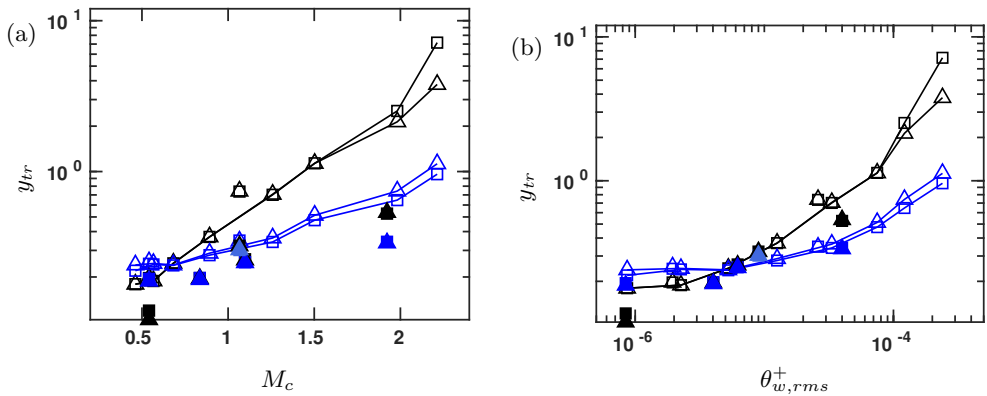
\begin{figure}
\centering
\includegraphics[width=.48\textwidth]{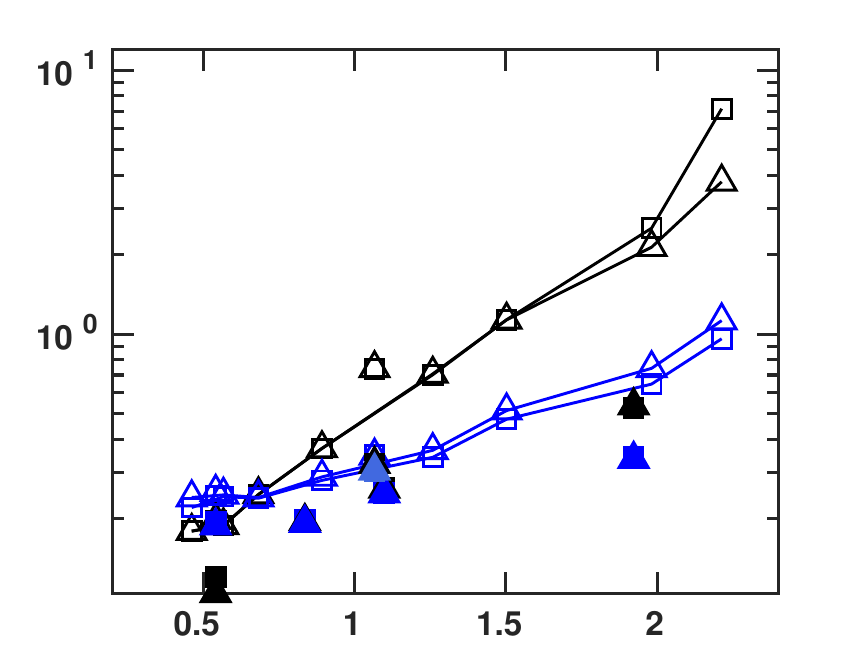}
\begin{picture}(0,0)
    \put(-198, 120){ (a)}
    \put(-190,60){\rotatebox{90}{ {$y_{tr}$}}}
    \put(-102, -9.5){ {$M_c$}}
\end{picture}
\includegraphics[width=.49\textwidth]{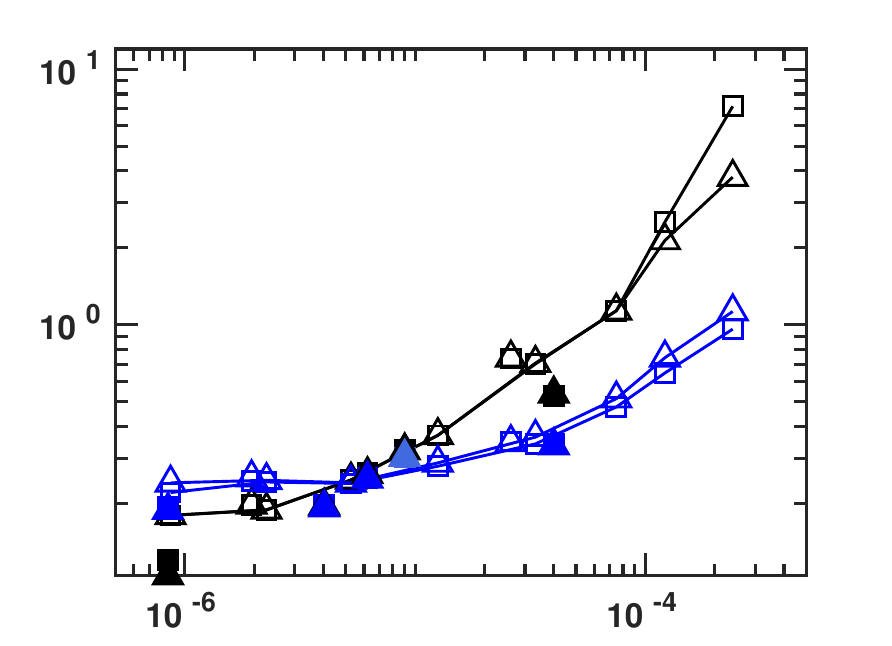}
\begin{picture}(0,0)
    \put(-202, 115){ (b)}
    \put(-195,60){\rotatebox{90}{{$y_{tr}$}}}
    \put(-105, -9.5){{$\dilwrms^+$}}
\end{picture}
\vspace {6mm}
\caption{Transition location of scaling exponents plotted versus
	(a) centerline Mach number (b) r.m.s dilatation at the wall.  
	Markers in all panels: ($\square$, $y^+$; $\triangle$, $y^*$)
	Black and Blue colored markers correspond to wall-normal Reynolds stress and
	shear Reynolds stress respectively for isothermal (empty markers), adiabatic (dark-filled markers) and pseudo-adiabatic (light-filled markers) cases.} 
\label{fig:ytr}
\end{figure}

Finally, in \rfig{ytr} we show
the wall-normal location where $R_{vv}$
and $R_{uv}$
transition from region R1 to region R2,
which is denoted as $y_{tr}$. 
Consistent with the results in \cite{BDB2022} we see 
in panel (a) 
that $y_{tr}$ moves away from the wall as $M_c$ is increased. 
However, we also observe clear WTBC effects.
In particular, we see that for adiabatic walls (dark-filled symbols) 
the transition moves closer to the wall compared to isothermal 
(empty symbols) and pseudo-adiabatic (light-filled symbols) walls.
For example, for high $M_c$, we see close to order-of-magnitude differences 
in $y_{tr}$ between isothermal and adiabatic cases for $R_{vv}$.
As before 
(\rfig{fgvtheta}(a)(b)), we can explore the suggestion in \cite{DJ2020} 
that a higher degree of universal behavior will be observed when dilatational
motions are used to scale statistics of interest. This is indeed supported
by the data in \rfig{ytr}(b) where we show $y_{tr}$ as a function of 
$\dilpwrms$.
Data for both $R_{vv}$ and $R_{uv}$ appear to be closer to exhibiting universal 
scaling (though not perfect) under this normalization. 

We can then conclude that by increasing the centerline Mach number 
or changing any other flow condition which results in enhancing 
dilatation levels at the wall, an enlarged region close
to the wall will develop where compressibility effects are significant.
This is also the region where Morkovin's hypothesis is found to be 
inadequate to collapse Reynolds stresses as shown before.

\section{Effects of thermal boundary conditions on temperature fluxes} \label{sec:temp}
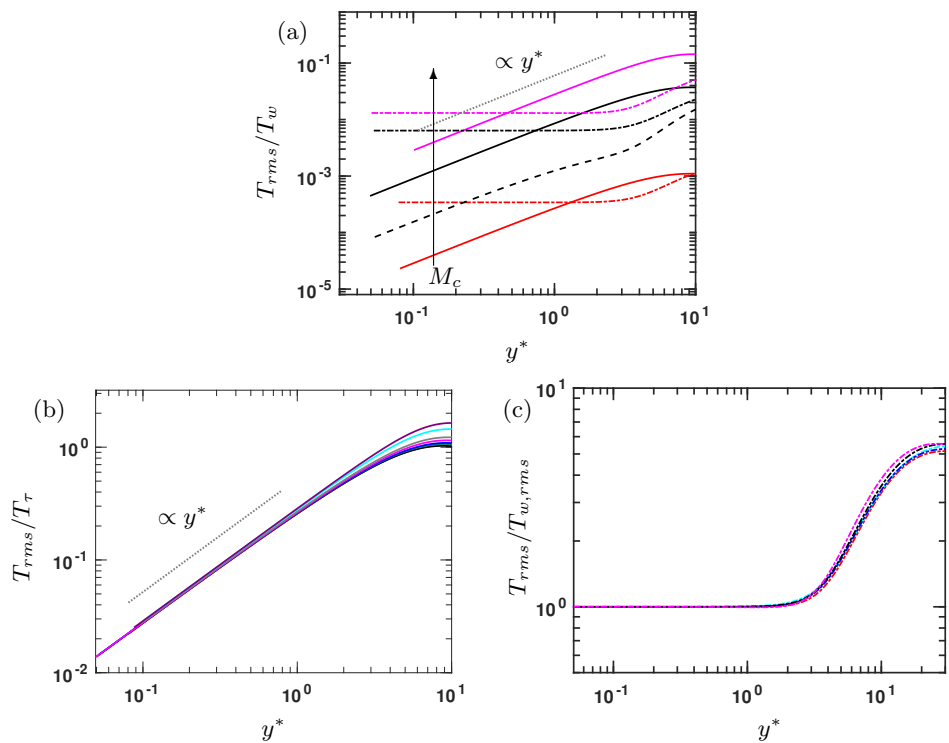
\begin{figure}
\centering
\includegraphics[width=.45\textwidth]{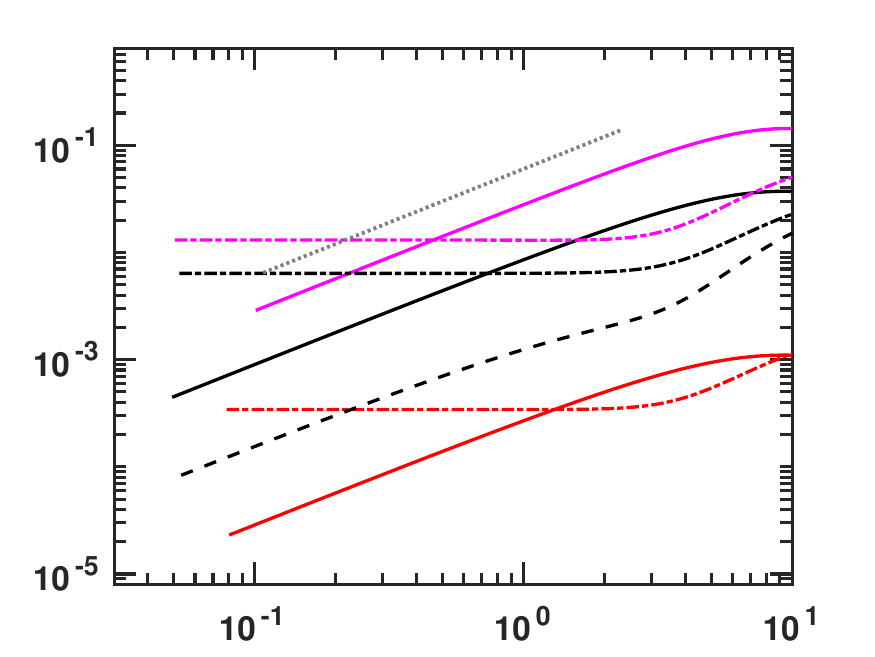}
\begin{picture}(0,0)
      \put(-180, 110){  (a)}
   \put(-185,50){\rotatebox{90}{{${T_{rms}}/T_w$}}}
 \put(-94, -9.5){  {$y^*$}}
  \put(-118,25){\vector(0,2){74}}
   \put(-120,18){$M_c$}
	\put(-95, 99.0){{$\propto y^*$}}
\end{picture}
\hskip 1.2 cm
\vskip 0.35 cm 
\includegraphics[width=.45\textwidth]{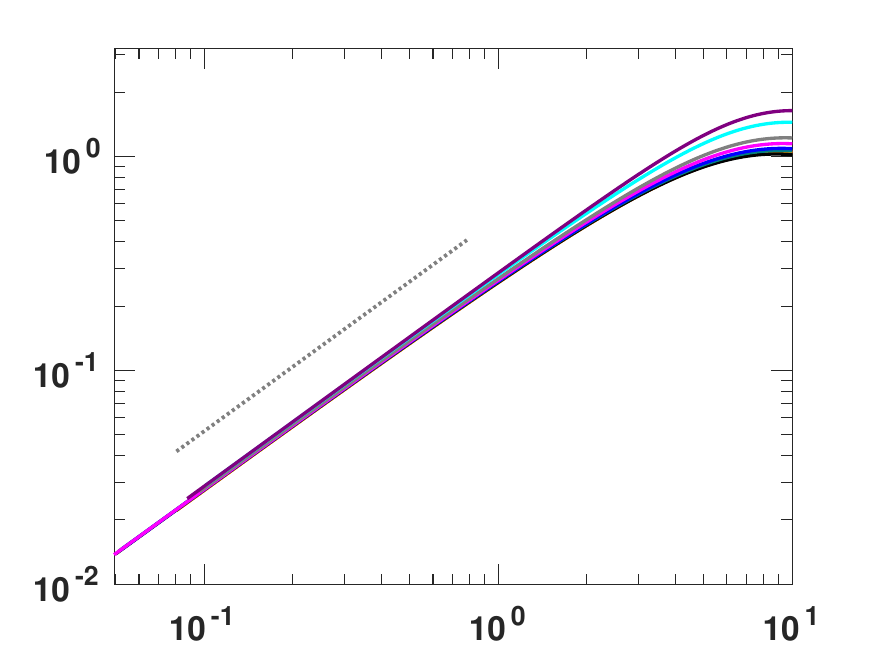}
\begin{picture}(0,0)
      \put(-180, 110){  (b)}
   \put(-185,45){\rotatebox{90}{ {${T_{rms}}/T_{\tau}$}}}
\put(-94, -9.5){  {$y^*$}}
	\put(-130, 70){{$\propto y^*$}}
\end{picture}
\includegraphics[width=0.47\textwidth]{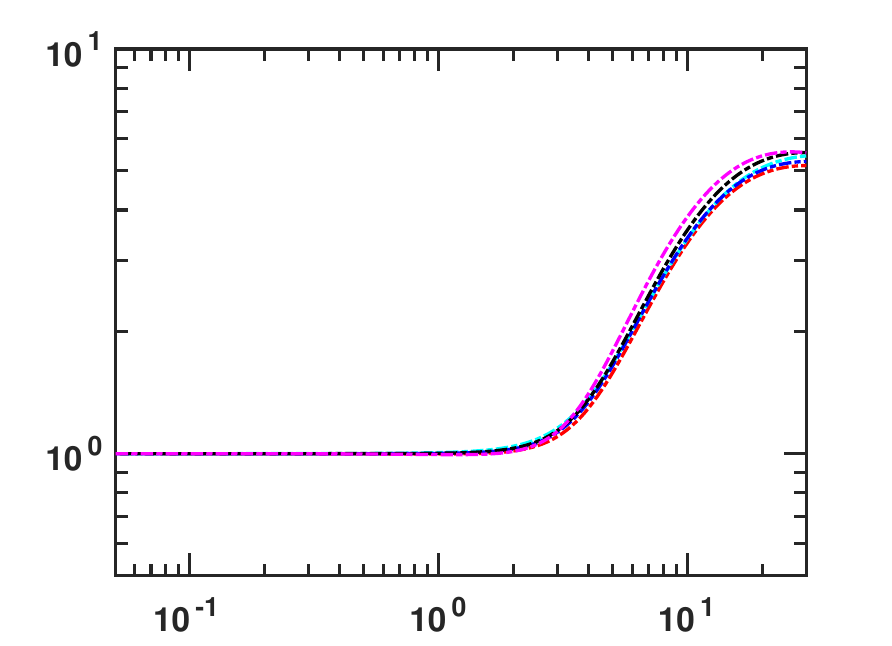}
\begin{picture}(0,0)
      \put(-190, 110){ (c)}
   \put(-185,40){\rotatebox{90}{ {${T_{rms}}/T_{w,rms}$}}}
\put(-94, -9.5){ {$y^*$}}
\end{picture}
\vskip 0.35 cm 
\caption{Root-mean-squared temperature fluctuations (a) 
normalized with wall temperature for isothermal   (---), adiabatic  
(-\,$\cdot$\,-) and pseudo-adiabatic (-\,-\,-) cases. 
(b) normalized with friction temperature for isothermal cases.
(b) normalized with r.m.s temperature at the adiabatic wall for adiabatic cases. Red, black and magenta correspond to $M_c \approx 0.23$, 
$M_c \approx 1.2$ and $M_c \approx 1.9$ respectively.}
\label{fig:tt}
\end{figure}

\subsection {Temperature fluxes}
The asymptotic behavior of temperature can be analyzed in a similar 
way to the velocity field 
by considering separately isothermal and adiabatic conditions.
The Taylor series expansion of temperature fluctuations is given by
\be
  T'=a_T + b_T y + c_T y^2 +\dots
\ee
For the isothermal case, temperature is fixed at the wall and 
one has $a_T=0$ (but $b_T\ne 0$).
For the adiabatic case, there are fluctuations at the wall but its normal gradient
vanishes,
in which case $b_T = 0$ (but $a_T\ne 0$). One would thus expect 
different near-wall asymptotic behavior based on thermal boundary conditions.
In \rfig{tt} (a), we plot the r.m.s.\ of temperature, $T_{rms}$, normalized 
by the mean wall temperature against 
$y^*$ for all cases.  
We find that the asymptotic behavior of $T_{rms}$ is qualitatively 
different for isothermal and 
adiabatic walls. Near the isothermal wall, $T_{rms}$ follows a 
power-law increase while the adiabatic cases are flat.
Similar asymptotic behavior was observed in incompressible and 
low-Mach number flows for isothermal and isoflux 
conditions \citep{TPLM2001,LSBH2009}. 
The asymptotic power-law scaling for isothermal cases
is equal to its theoretical asymptote ($\gamma_{T} = 1$)
for all $M_c$.
For adiabatic cases, the profile is constant for most of the 
viscous sublayer (until $y^* \approx 2$) and that constant
increases with increase in the $M_c$.
Interestingly, the pseudo-adiabatic case (black dashed line), 
which has been extensively used in the
literature to model adiabatic walls, exhibits an isothermal-like power-law
behavior close to the wall. 

\begin{figure}
\includegraphics[width=.45\textwidth]{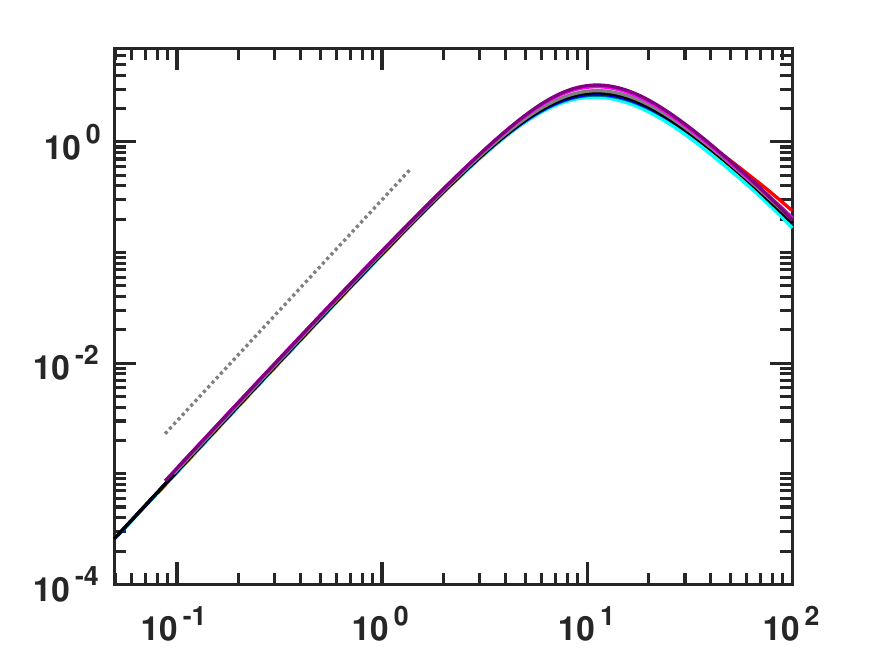}
\begin{picture}(0,0)
   \put(-185,30){\rotatebox{90}{{$\ol{\rho} \wt{u''T''}/(\rho_w u_\stau T_\stau)$}}}
	\put(-140, 75.0){{$\propto y^2$}}
    \put(-185, 110){ (a)}
\put(-90, -9.5){ {$y^*$}}
\end{picture}
\hskip 0.4 cm
\includegraphics[width=.45\textwidth]{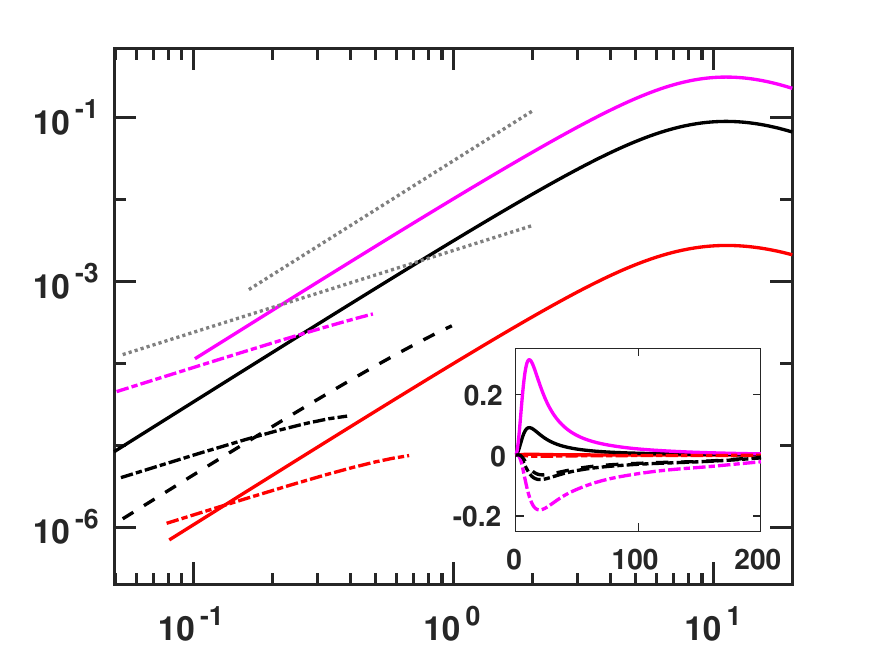}
\begin{picture}(0,0)
        \put(-185, 110){ (b)}
   \put(-185,30){\rotatebox{90}{{$\ol{\rho} \wt{u''T''}/(\rho_w u_\stau T_w)$}}}
 \put(-90, -9.5){  {$y^*$}}
	\put(-120, 90){{$\propto y^2$}}
	\put(-148, 70.0){{$\propto y$}}
   \end{picture}
\vspace {6mm}
  \caption{Streamwise turbulent heat flux close to (a) isothermal walls normalized 
by friction temperature. (b) isothermal (---), adiabatic (-\,$\cdot$\,-) and pseudo-adiabatic 
(-\,-\,-) cases normalized by their respective wall temperature. Inset contains the same data in linear scales upto $y^* \approx 300$. Red, black and magenta correspond to $M_c \approx 0.23$, 
$M_c \approx 1.2$ and $M_c \approx 1.9$ respectively.} 
\label{fig:ut}
\end{figure}
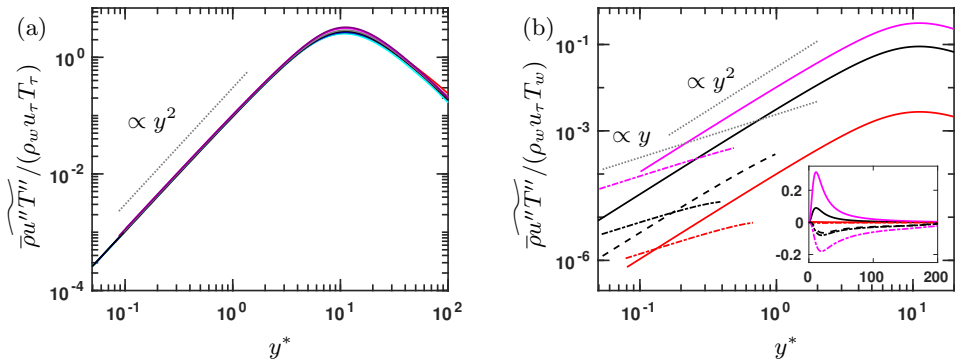

An alternative normalization for temperature, in analogy with the 
Reynolds stresses, is through the so-called 
friction temperature, 
$T_{\tau} \equiv -\kappa (\partial{\ol{T}}/\partial{y})_w/
\ol{\rho_w} c_p u_\stau$, where $\kappa$ is the thermal conductivity.
It is clear, however, that this normalization can only be applied 
to isothermal walls since adiabatic (and pseudo-adiabatic) walls
present zero conductive heat transfer to the wall 
($\partial \ol{T}/\partial y|_w=0$).
In \rfig{tt} (b), we show all isothermal cases which do, in fact, collapse 
in the near-wall region following its asymptotic scaling of $\sim y^*$.
A collapse of adiabatic cases is also obtained when $\Trms$
is normalized with 
their respective wall values ($T_{w,rms}$) as seen in \rfig{tt} (c).
Since $T_{w,rms}=0$ for isothermal cases, it is clear that neither normalization
provides universal scaling across different WTBCs.

\subsubsection{Streamwise turbulent heat flux}
The streamwise component of the turbulent heat-flux ($R_{uT}$) is an important quantity in
wall-bounded flows which needs to be correctly modeled in order
to make accurate predictions. 
In fact, this heat-flux component has been found to be even larger than the wall-normal turbulent heat flux (\cite{aiaa_HNLMB2020}). 
Current Boussinesq or constant $Pr_T$ based RANS models, however,
cannot capture its behavior accurately \citep{bowersox2009,HBD2019,BMB2022}. 

In \rfig{ut} (a) we show the density-scaled streamwise turbulent heat flux, 
$\ol{\rho} \wt{u''T''}/(\rho_w u_\stau T_\stau)$ 
in the near-wall region of an 
isothermal wall for different Mach numbers against $y^*$. 
We observe very good collapse for all Mach numbers along the theoretical 
asymptotic power law given by $\gamma_{uT} = 2$ (\rtab{gamma}).
The adiabatic and pseudo-adiabatic cases are included in \rfig{ut} (b)
(normalized with wall temperature)
along with the isothermal cases for comparison.
The temperature fluctuations at the wall for
adiabatic cases result in 
$\gamma_{uT} = 1$ and  again conforms to the theoretical
behavior. 
Following $T_{rms}$, we 
observe that power-law behavior for pseudo-adiabatic streamwise heat flux follows isothermal-like behavior 
and thus, also matches with the isothermal theoretical exponent. 
The streamwise heat flux becomes negative for $y^*\gtrsim 1 $ 
 in adiabatic and pseudo-adiabatic cases 
and thus can not be shown in logarithmic scales. Again, this indicates that fine resolution close to the wall
is required to capture correct near-wall asymptotic behavior.
In inset of \rfig{ut} (b), we also include streamwise heat-flux along the 
channel in linear scales. Similar to $T_{rms}$, we find that any scaling, 
either by normalization using $T_w$ or $\ol{T}$ (not shown here), 
$R_{uT}$ do not collapse for different $M_c$ and WTBCs in high-speed regime.

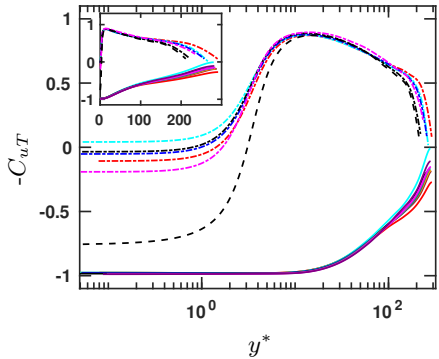
\begin{figure} 
\begin{center} 
\includegraphics[width=.45\textwidth]{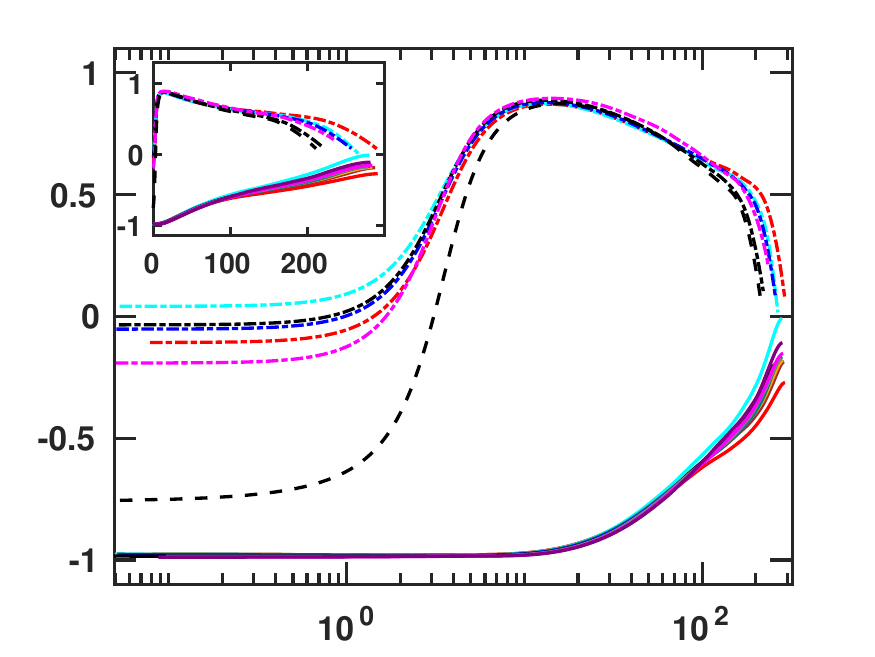}
\begin{picture}(0,0)
    \put(-180,50){\rotatebox{90}{  {-$C_{uT}$}}}
    \put(-90, -9.5){\small {$y^*$}}
\end{picture}
\end{center}
\vskip 0.4 cm
   \caption{Correlation coefficient for $R_{uT}$ for isothermal 
        (---), adiabatic (-\,$\cdot$\,-) and pseudo-adiabatic (-\,-\,-) cases.
     Colors as in \rtab{dns1}. Inset contains the same data in linear scales. }
\label{fig:cut}
\end{figure}

Similar to the case of $R_{uv}$ discussed above, the near-wall asymptotic behavior
of $R_{uT}$ will depend not only
on the scaling of the r.m.s. of the two variables involved in the flux, 
but also on their cross-correlation.
The excellent agreement seen for $\gamma_{uT}$ for all cases
with their respective theoretical scaling, then, implies that the 
correlation coefficient, $\cut$ does not 
vary in $y$ in this region and is evident in \rfig{cut}. 
For $y^{*} \lesssim 1$, $\cut$ is constant with $y^{*}$ for all $M_c$ and WTBCs.
However, we see interesting differences between different WTBCs. 
First, the absolute value of $\cut$ is minimum near 
the wall for adiabatic cases while for isothermal cases, the absolute value of $\cut$ 
is maximum close to the wall. Some $M_c$ effects can be observed 
for adiabatic cases close to the wall while -$\cut$ for different 
$M_c$ collapses to a constant value of $-1$ near isothermal walls. For adiabatic cases, the 
decorrelation decreases on moving away from the wall until 
$y^* \approx 15$, while $\cut$ for isothermal cases remains constant in this 
region, with -$\cut \approx -1$. Interestingly, the pseudo-adiabatic case 
resembles isothermal-like behavior near the wall and adiabatic-like behavior beyond $y^* \approx 10$. At further distance, $y^* \gtrsim 15$, 
decorrelation increases for all WTBCs with isothermal case maintaining 
a positive correlation, while adiabatic and pseudo-adiabatic maintaining  
negative correlations ($\cut$). The correlation approaches zero as one 
moves towards the centerline. For $y^* \gtrsim 100$, Mach number effects can be
seen for 
isothermal and adiabatic cases. Another interesting observation from \rfig{cut}
is that 
$\cut$ for adiabatic and pseudo-adiabatic cases, as shown in the inset 
of \rfig{cut} resemble $\cut$ profile for a flat-plate boundary layer 
(\cite{DBM2010}) with isothermal or pseudo-adiabatic walls. 

\subsubsection{Wall-normal turbulent heat flux}
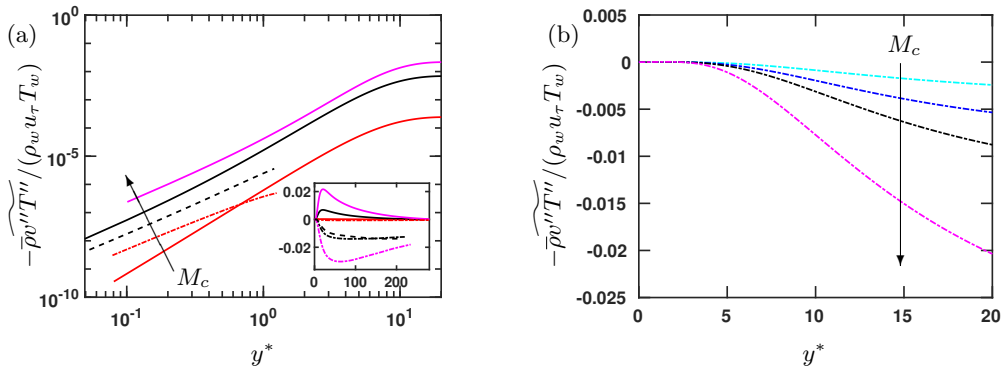
\begin{figure}
\centering
\includegraphics[width=.45\textwidth]{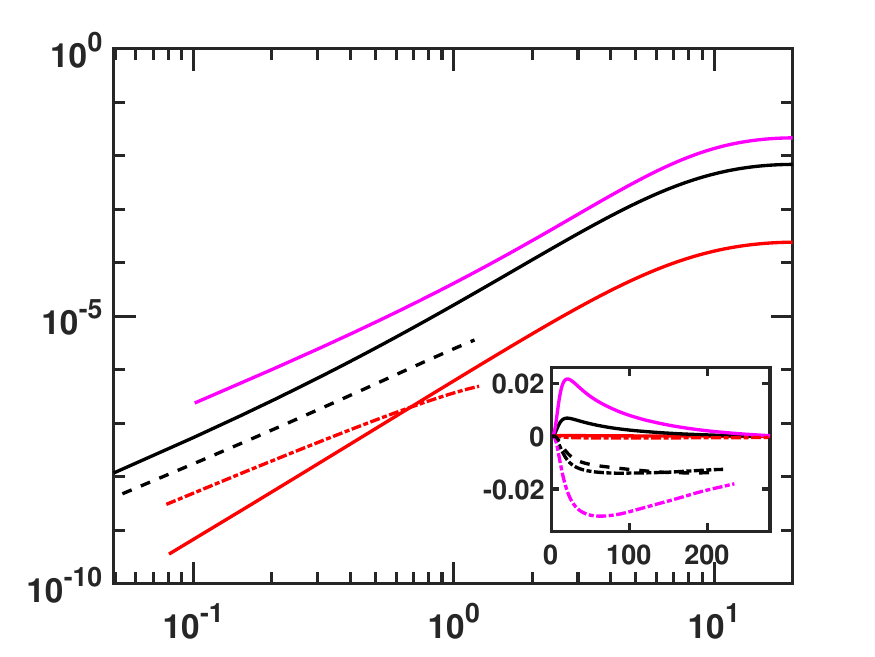}
\begin{picture}(0,0)
      \put(-186, 110){  (a)}
   \put(-182,22){\rotatebox{90}{ {$-\ol{\rho} \wt{v''T''}/(\rho_w u_\stau T_w)$}}}
 \put(-94, -9.5){  {$y^*$}}
   \put(-120,24){\vector(-1,2){18}}
   \put(-122,19){ $M_c$}
\end{picture}
\hskip 1.0 cm
\includegraphics[width=.45\textwidth]{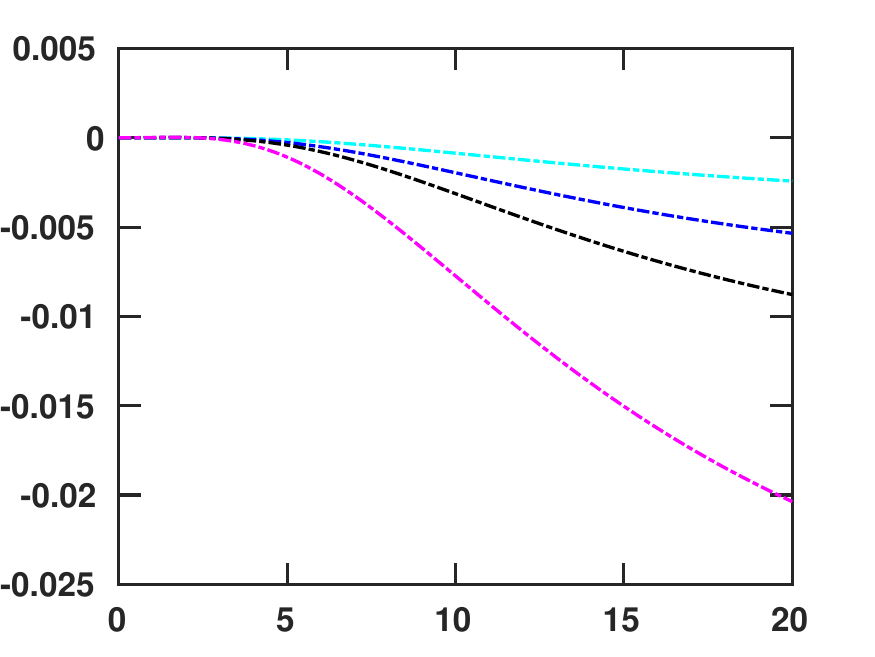}
\begin{picture}(0,0)
      \put(-190, 110){  (b)}
   \put(-192,22){\rotatebox{90}{ {$-\ol{\rho} \wt{v''T''}/(\rho_w u_\stau T_w)$}}}
 \put(-94, -9.5){  {$y^*$}}
  \put(-54,102){\vector(0,-2){76}}
   \put(-62,106){  $M_c$}
\end{picture}
\vskip 0.35 cm 
\caption{Wall-normal turbulent heat flux close to (a) isothermal, 
        (---), adiabatic (-\,$\cdot$\,-) and pseudo-adiabatic (-\,-\,-) walls
in logarithmic scale. Inset contains the same data in linear scales upto $y^* \approx 300$.
	Red, black and magenta correspond to $M_c \approx 0.23$, 
$M_c \approx 1.2$ and $M_c \approx 1.9$ respectively.
(b) adiabatic walls in linear scales. Colors as in \rtab{dns1}.} 
\label{fig:vt}
\end{figure}

\begin{figure}
\centering
\includegraphics[width=.45\textwidth]{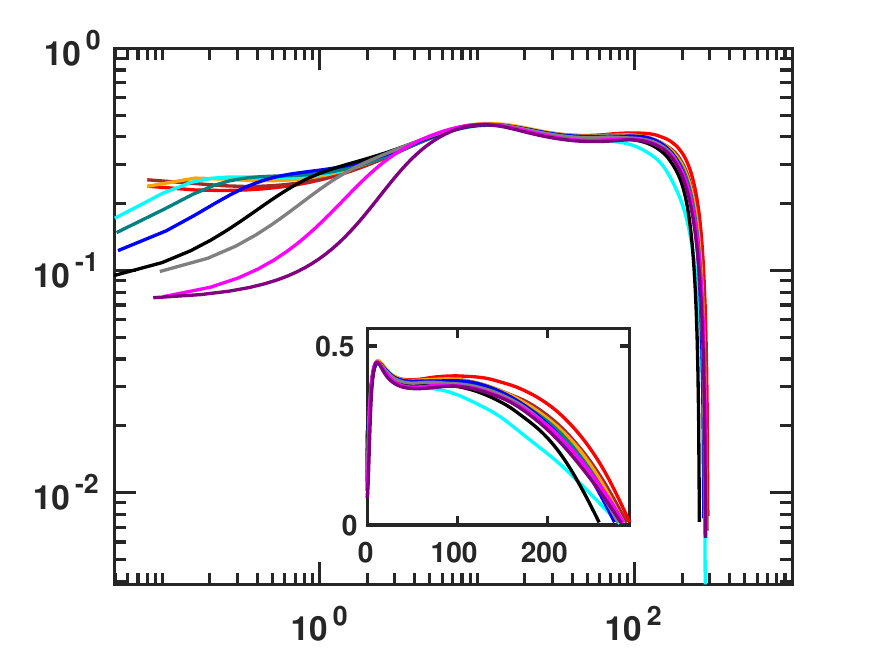}
\begin{picture}(0,0)
    \put(-186, 108){\small (a)}
    \put(-185,55){\rotatebox{90}{ {-$C_{vT}$}}}
    \put(-88, -3.5){\small {$y^*$}}
   \put(-145,102){\vector(1,-2){20}}
   \put(-149,105){$M_c$}
\end{picture}
\includegraphics[width=.45\textwidth]{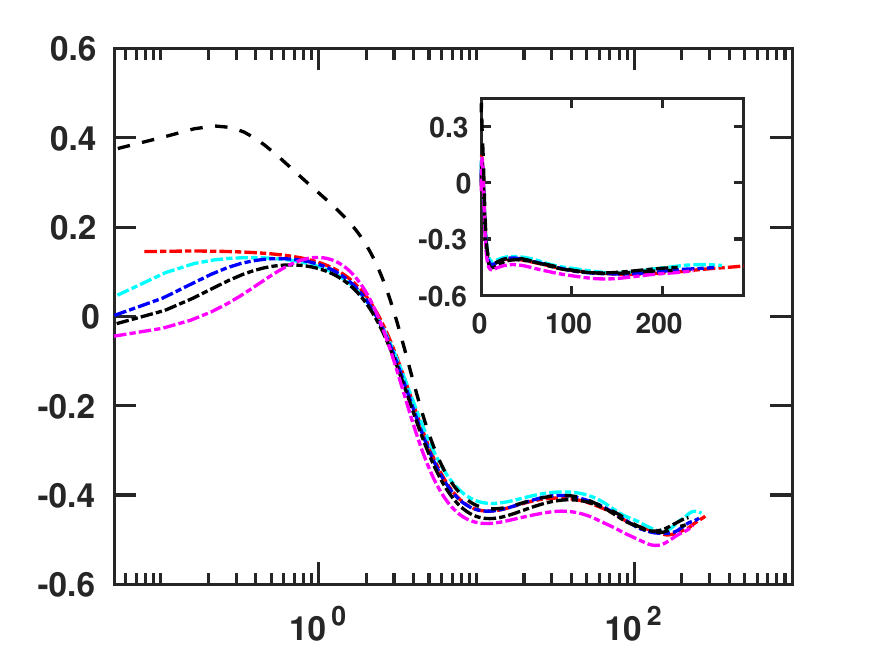}
\begin{picture}(0,0)
    \put(-186, 108){\small (b)}
    \put(-185,55){\rotatebox{90}{  {-$C_{vT}$}}}
    \put(-88, -3.5){\small {$y^*$}}
    \put(-140,84){\vector(1,-2){12}}
   \put(-148,88){{$M_c$}}
\end{picture}
	\caption{ Correlation coefficient for 
$R_{vT}$ near (a) isothermal walls 
        (b) adiabatic  (-\,$\cdot$\,-) and pseudo-adiabatic 
(-\,-\,-) walls.
Inset has the same data in linear scales. Colors as in \rtab{dns1}.}
\label{fig:cvt}
\end{figure}
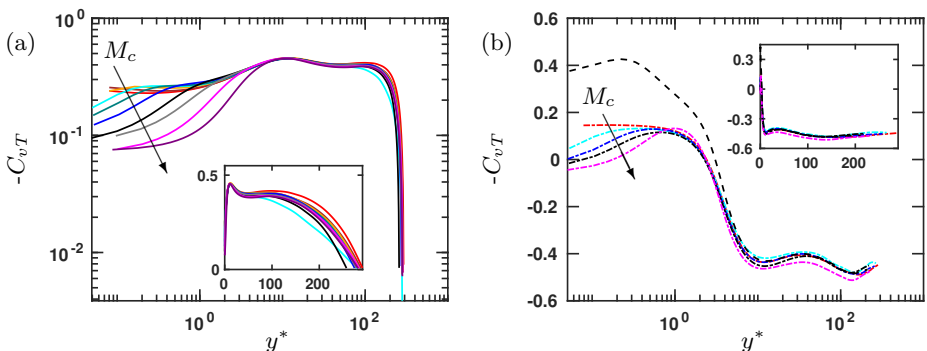

In \rfig{vt} (a), we plot the density averaged wall-normal turbulent heat flux ($R_{vT}$), 
$\ol{\rho} \wt{v''T''}/(\rho_w u_\stau T_w)$ close to the wall for 
isothermal cases (solid lines) with $M_c \approx {0.23,1.2,1.9} $, 
low Mach adiabatic case with $M_c \approx 0.23$ and pseudo-adiabatic 
case with $M_c \approx 1.2$.  
It can be seen that close to the isothermal wall, 
a Mach number dependent power-law exists for the wall-normal 
turbulent heat-flux. A detailed study of asymptotic 
power-law for wall-normal turbulent heat flux close to 
isothermal walls was performed in \cite{BDB2022},
where power-law 
exponents were observed to transition from its theoretical low Mach to high 
Mach asymptotes. This transition was found to be similar to that of $\gamma_{uv}$. 
The asymptotic behavior of heat-flux close to the 
pseudo-adiabatic wall exhibits a power-law behavior 
with $\gamma_{vT} \approx 2.1$ and matches closely to the 
theoretical limit of isothermal asymptotic power law. 
This is in line with the behavior of all other statistics 
close to the pseudo-adiabatic wall which behave like 
those in isothermal cases. For Mach number in the 
near-incompressible range $M_c=0.23$, a power-law behavior with 
exponent equals its theoretical value is observed close to the 
adiabatic wall. 
Similar to $R_{uT}$, $R_{vT}$ for this adiabatic case ($M_c=0.23$), and 
pseudo-adiabatic case changes sign moving away from the wall 
and therefore can not be shown in logarithmic scales. For adiabatic cases 
with $M_c > 0.23$, we find that a well-defined 
power-law behavior is not observed close to the wall and hence the 
data is plotted in 
linear scales in \rfig{vt} (b).  
Similar to $T_{rms}$ and $R_{uT}$, we find that any 
scaling, either by normalization using $T_w$ or $\ol{T}$ (not shown here), 
$R_{vT}$ do not collapse for different $M_c$ and WTBC in high-speed regimes. 

Finally, we plot $-C_{vT}$ as a function of $y^*$
in \rfig {cvt} (a) and (b). For isothermal cases as shown in \rfig{cvt}(a),
$C_{vT}$ closely resembles $C_{uv}$ as shown in \rfig{cuv} indicating that 
increasing $M_c$ has similar effects on $C_{vT}$
as were observed for $C_{uv}$. In the inset of \rfig{cvt}(a), 
we plot $C_{vT}$ 
in linear scales where moving towards the centerline ($y^* \gtrsim 100$), 
some Mach number effects can be observed.
 For adiabatic walls, as is shown in \rfig{cvt} (b), a trend with the Mach number 
close to the wall is observed for $C_{vT}$. 
On moving away from the wall, the decorrelation between $v'$ and $T'$ decreases.  
Furthermore, the effect of mixed boundary condition
can be observed close to the channel centerline where $C_{vT}$ does not vanish. This is because of the
finite mean temperature gradients at the channel half-width resulting in the non-zero 
wall-normal heat flux at $h$. On comparing \rfig{cvt} (a) and (b),
we observe that $C_{vT}$ assumes opposite signs 
for isothermal and adiabatic cases in regions away from the wall.
Like the previous observation for other statistics close to the wall,
pseudo adiabatic exhibits isothermal-like near-wall behavior but 
follows adiabatic in regions away from the wall.

\section{Conclusions}
The asymptotic behavior of turbulent stresses and turbulent heat fluxes close
to the wall were investigated using a large DNS database of 
turbulent channel flows with centerline Mach numbers spanning 
from 0.23 to 2.22.
The dataset comprises of simulations with three different 
wall thermal boundary conditions (WTBC), namely
isothermal, adiabatic and pseudo-adiabatic.
A distinguishing feature of the present DNS is the near-wall resolution which 
is much finer than those typically found in the literature. 
We show this is essential to capture near-wall behavior for 
different flow and wall boundary conditions. 

Turbulent stresses containing wall-normal velocity  
component 
do not exhibit a universal behavior close to the wall when normalized
using either wall or semi-local units.
Interestingly, some statistics behave differently for different 
WTBCs while others behave similarly.
Similarities include Mach number effects on statistics close 
to the wall for isothermal and adiabatic cases. 
In both cases, turbulent stresses exhibited
asymptotic power-law behavior in the near-wall region (which we call R1) for
all Mach numbers and WTBCs. With increase in Mach number, smooth transition of
asymptotic power-law exponents from the solenoidal limit to the high-speed
limit was observed. Consistent with previous 
findings, a second scaling regime (R2) with a steeper exponent and a
weaker Mach number dependence beyond R1 was observed. The transition
location between R1 and R2 was dependent on Mach number.

A notable difference between cases with different WTBCs is the change in
power-law exponents for turbulent stresses with changing WTBC at high 
Mach numbers. This effect is stronger for $R_{uv}$ than for $R_{vv}$. 
In general, $R_{uv}$ was found to be more
sensitive to changes in $M_c$ or WTBC. This was linked to a
decorrelation between $u'$ and $v'$ when $M_c$ is increased or
when the WTBC changes from isothermal to adiabatic.

Inspired by a recent proposal based on homogeneous flows, 
we found that universality can be indeed recovered if dilatational
motions are incorporated as a governing parameter regardless of 
the mechanism that generated them.
In particular, asymptotic power-law exponents and the transition location between the two scaling regimes R1 and R2 
do collapse on a universal curve which depends uniquely 
on $\dilwrms$, the r.m.s.\ of dilatation 
at the wall. If one uses the (perhaps more intuitive)
centerline Mach number one can clearly see differences in  
exponents and transition location for different WTBCs.
This clearly supports the idea that dilatational levels,
regardless of how they are generated, provide the
appropriate scaling parameter for near-wall behavior
at high speeds furthering the idea of 
some universality of statistics in compressible wall-bounded flows. 
This also support the previously found conclusion that 
Morkovin's hypothesis does not take into consideration all 
the effects associated with compressibility at higher Mach numbers.

We also investigated statistics of temperature fluctuations, 
wall-normal and streamwise turbulent heat fluxes for varying $M_c$ and WTBC.
For isothermal cases we 
found that $T_{rms}$ follows a power-law behavior predicted
by the analytical form of its Taylor expansion.
For adiabatic cases, 
on the other hand,
$T_{rms}$ remains constant in the viscous sublayer
followed by an almost universal increase with $y^*$.

The streamwise heat flux, $R_{uT}$, exhibits a power-law
behavior close to the wall with exponents given by theoretical predictions 
for both isothermal and adiabatic cases.
In general, it was found that  
temperature statistics ($T_{rms}$, $R_{uT}$) can be collapsed separately
for isothermal and adiabatic cases
by normalizing temperature with $T_{\tau}$ and $T_{w,rms}$,
respectively.
However, no general scaling laws were found that could collapse
statistics containing temperature fluctuations for
both WTBCs.

As with Reynolds stresses,
the wall-normal turbulent heat flux ($R_{vT}$)
for isothermal cases exhibits power-law behavior
with exponents that depend on $M_c$.
A well-defined power-law behavior
cannot be unambiguously identified for adiabatic cases with
$M_c > 0.23$.
Pseudo-adiabatic walls, which are often used to mimic an adiabatic wall 
by imposing an isothermal condition at the adiabatic temperature,
displayed isothermal-like behavior close to the wall as $M_c$ increases.
A rich interplay between Mach number and WTBC effects was
observed for 
correlation coefficients between $v'$ and $T'$, and between $u'$ and $T'$
indicating a complex dynamics between velocity and temperature fluctuations. 

Mach number effects were observed in the viscous sublayer for
the correlation between $v'$ and $T'$ for all WTBCs,
but only in the adiabatic case for $u'$ and $T'$.
The strong WTBC effect is evident by the fact that
these correlations possess different signs in most of the  
region across the channel.
In these regions, $v'$ and $T'$ are negatively
correlated for isothermal walls  
while they are positively correlated for adiabatic
cases. 
In contrast, $u'$ and $T'$ are positively correlated for
isothermal walls while negatively correlated for adiabatic cases. 
Moreover, in the
region close to the wall, the magnitude of these $u'$ and $T'$ correlations are very different
for isothermal and adiabatic cases, with the former being much stronger 
than the latter.
Similar to all other statistics,
pseudo-adiabatic case exhibits isothermal-like near-wall behavior but resembles
the adiabatic profile away from the wall.

We close by pointing out that, overall, 
Morkovin's hypothesis and semi-local normalizations do not collapse
data for all the flow and boundary conditions. Universal scaling
laws for wall-bounded compressible flows, thus, 
requires more general scaling laws.

\ \\

Acknowledgments.
The authors acknowledge support from (1) the National Science Foundation (Grant No.~1605914), (2) DoD Vannevar Bush Faculty Fellows (ONR Grant No. N00014-18-1-3020), and (3) the Extreme Science and Engineering Discovery Environment (XSEDE) for computational resources. The opinions, findings, views, conclusions, or recommendations contained herein are those of the authors and should not be interpreted as necessarily representing the official policies or endorsements, either expressed or implied, of the U.S. Government.

Competing interests: The authors declare none.

\bibliographystyle{jfm}

\end{document}